\documentclass[prb,twocolumn,showpacs,floatfix]{revtex4}

\usepackage{amsmath}
\usepackage{amssymb}
\usepackage{tikz}
\usetikzlibrary{arrows}
\usepackage{dcolumn}

\providecommand{\eqref}[1]{(\ref{#1})}

\newcommand{\abs}[1]{\left\vert#1\right\vert}

\newcommand{\eps}{\varepsilon}

\newcommand{\phdag}{{\phantom{\dag}}}

\newcommand{\conj}{{*}}
\newcommand{\phconj}{{\phantom{\conj}}}

\newcommand{\NitrogenPositiveIonTermSymbol}{\ensuremath{^2\Pi_u}}
\newcommand{\NitrogenPositiveIon}{\ensuremath{\mathrm{N_2^+}(\NitrogenPositiveIonTermSymbol)}}

\newcommand{\NitrogenGroundStateTermSymbol}{\ensuremath{^1\Sigma_g^+}}
\newcommand{\NitrogenGroundState}{\ensuremath{\mathrm{N_2}(\NitrogenGroundStateTermSymbol)}}

\newcommand{\NitrogenDominantMetastableStateTermSymbol}{\ensuremath{^3\Sigma_u^+}}
\newcommand{\NitrogenDominantMetastableState}{\ensuremath{\mathrm{N_2}(\NitrogenDominantMetastableStateTermSymbol)}}

\newcommand{\NitrogenNegativeIonResonanceTermSymbol}{\ensuremath{^2\Pi_g}}
\newcommand{\NitrogenNegativeIonResonance}{\ensuremath{\mathrm{N_2^-}(\NitrogenNegativeIonResonanceTermSymbol)}}

\newcommand{\RCTCaptureME}{V_{\vec{k}}}
\newcommand{\RCTSIReleaseME}{V_{\vec{q}}}
\newcommand{\DPDME}{V_{\vec{k}\vec{q}}}

\begin{document}

%\title{Slave-boson approach to the coupling of resonant and Auger processes in the surface de-excitation of 
%        metastable nitrogen molecules}
\title{Pseudo-particle approach for charge-transferring molecule-surface collisions}

\author{Johannes Marbach}
\email[]{marbach@physik.uni-greifswald.de}
\author{Franz Xaver Bronold}
\author{Holger Fehske}
%\homepage[]{Your web page}
%\thanks{}
\affiliation{Institut f\"ur Physik, Ernst-Moritz-Arndt-Universit\"at Greifswald, 17489 Greifswald, Germany}

\date{\today}

\begin{abstract}
Based on a semi-empirical generalized Anderson-Newns model we construct a pseudo-particle description 
for electron emission due to de-excitation of metastable molecules at surfaces. The pseudo-particle
approach allows us to treat resonant charge-transfer and Auger processes on an equal footing, as it 
is necessary when both channels are open. This is for instance the case when a metastable
\NitrogenDominantMetastableState\ molecule hits a diamond surface. Using non-equilibrium Green functions and 
physically motivated approximations to the self-energies of the Dyson equations we derive a system of 
rate equations for the probabilities with which the metastable \NitrogenDominantMetastableState\ molecule, 
the molecular ground state \NitrogenGroundState , and the negative ion \NitrogenNegativeIonResonance\ can 
be found in the course of the scattering event. From the rate equations we also obtain the spectrum of the 
emitted electron and the secondary electron emission coefficient. Our numerical results indicate  
the resonant tunneling process undermining the source of the Auger channel which therefore contributes 
only a few percent to the secondary electron emission. 
\end{abstract}


\maketitle

\section{Introduction}

Charge exchange processes during atom-surface or molecule-surface collisions have been the subject 
of intense scientific research during the last decades~\cite{Winter2007,Rabalais1994}. This type 
of surface reactions is of fundamental interest. It represents a quantum-impurity problem where a 
finite many-body system with discrete quantum states couples to an extended  
system with a continuum of states which essentially acts as a reservoir for electrons. Under 
appropriate conditions~\cite{He2010,Merino1998Roomtemperature,Shao1994Manybody} such an arrangement
gives for instance rise to the Kondo effect~\cite{KondoEffect1993} originally found in metals containing 
magnetic impurities or to Coulomb blockades as it is discussed in nanostructures~\cite{CoulombBlockade1992}. 

Besides of being a particular realization of a quantum impurity problem, atom/molecule-surface 
collisions are also of technological interest, especially in the field of bounded low-temperature 
plasmas, where this type of surface collisions is the main supplier of secondary electrons which in turn 
strongly affect the overall charge balance of the discharge~\cite{Lieberman2005}. In dielectric barrier 
discharges, for instance, secondary electron emission determines whether the discharge operates in a 
filamentary or a diffuse mode~\cite{Brandenburg2005,Massines2003}. Only the latter mode is useful for 
surface modification. Controlling the yield with which secondary electrons are produced is thus of great 
practical interest. This applies even more so to micro-discharges~\cite{Becker2006} where the continuing 
miniaturization gives charge-transferring surface reactions more and more influence on the properties of 
the discharge. 

Depending on the projectile and the target, secondary electron emission usually occurs either in the 
course of a resonant tunneling process or an Auger transition. In some situations, however, both 
transitions may be energetically allowed and hence contribute to the yield with which electrons are 
released. The interplay of the two reaction channels has therefore been studied in the 
past~\cite{Snowdon1987,Zimny1991,Onufriev1996Memory,Keller1998,Lorente1998Unified,Alvarez1998Auger,Goldberg1999,Pepino2002,Wang2001,Garcia2003Interference}.
Starting with the work of \citeauthor{Alvarez1998Auger}~\cite{Alvarez1998Auger} a detailed theoretical 
analysis of the interference of Auger and resonant tunneling processes has been given by Goldberg and 
coworkers~\cite{Goldberg1999,Wang2001,Garcia2003Interference}. Their results for $\mathrm{H}^+$ and $\mathrm{He}^+$ 
indicate the Auger channel to be active only close to the surface whereas the resonant channel is already 
efficient at rather large projectile-surface distances. When both channels are coupled together the dynamics 
of the system is hence controlled by the resonant channel as it destroys the initial species before the Auger 
channel can become operative. The Auger channel is therefore strongly suppressed in the coupled system albeit
the individual efficiencies of the reaction channels are comparable. Onufriev and Marston~\cite{Onufriev1996Memory} 
also investigated the interplay of tunneling and Auger processes for the particular case of $\mathrm{Li}$(2p) 
atoms de-exciting on a metallic surface. Using a sophisticated many-body theoretical description of the 
scattering process they concluded that depending on the model parameters the two de-excitation channels 
interfere either constructively or destructively.

Whereas the previous studies focused on atomic projectiles we will investigate in the following the interplay 
of Auger and resonant tunneling processes for a molecular projectile. More precisely, we will analyze 
how these two processes affect secondary electron emission due to de-excitation of metastable 
molecules. Neutralization of molecular ions~\cite{Heiland94,Imke1986Resonant,Imke1986Theory} will not be 
discussed. A particularly interesting case is the de-excitation of metastable \NitrogenDominantMetastableState\ 
because this molecule de-excites in two primary reaction channels~\cite{Stracke1998Formation,Lorente1999N2}. 
On the one hand, there is the two-step resonant charge transfer (RCT) reaction
\begin{equation}
  e_{\vec{k}}^- + \NitrogenDominantMetastableState \xrightarrow{\text{RCT}} \NitrogenNegativeIonResonance
\xrightarrow{\text{RCT}} \NitrogenGroundState + e_{\vec{q}}^- \;,
  \label{eq react rct}
\end{equation}
where~$e_{\vec{k}}^-$ and~$e_{\vec{q}}^-$ denote an electron within the surface and a free electron, respectively. In
this process the metastable~\NitrogenDominantMetastableState\ molecule first resonantly captures an electron from the
surface to form the intermediate negative ion shape resonance \NitrogenNegativeIonResonance\ which then decays into
the ground state~\NitrogenGroundState\ by resonantly emitting an electron. The decay of the negative ion can be either
due to the ion's natural life time or due to the ion surface interaction. In addition to the RCT channel, there exists
an Auger de-excitation reaction also known as Penning de-excitation
\begin{equation}
  e_{\vec{k}}^- + \NitrogenDominantMetastableState \xrightarrow{\text{Auger}} \NitrogenGroundState + e_{\vec{q}}^- \;.
  \label{eq react auger}
\end{equation}
Here the molecule non-resonantly captures a surface electron and simultaneously releases another electron. Depending
on the surface band structure both processes may be possible at the same time. This is for instance the case 
for diamond as it possesses a rather wide valence band. 

In our previous work~\cite{Marbach2011Auger,Marbach2012Resonant} we investigated the two reaction channels 
separately by a quantum-kinetic approach and a rate equation technique. The molecule was in both cases 
treated as a semi-empirical two-level system corresponding to the $2\pi_u$ and $2\pi_g$ molecular 
orbitals, which are the two molecular orbitals whose occupancies change during the de-excitation 
process. Spectator electrons not involved in the processes were neglected. 
%In particular the three spectator electrons in the 
%$2\pi_u$ level were assumed to be frozen in, restricting thus the occupancy of this level to one.
Depending on the process the two levels denoted the upper and lower ionization levels of either the 
metastable molecule (Auger de-excitation) or the negative ion (resonant charge-transfer). 
The advantage of a semi-empirical model is that it is based on a few parameters which are relatively 
easy accessible, either experimentally or theoretically. The difference of the two parameter sets, 
which arises from intra-molecular Coulomb correlations not included in the model, is not a problem as 
long as the two processes are treated separately. A simultaneous treatment of them requires
however a way to implement both parameterizations within a single Hamiltonian.

A way to overcome the problem would be of course to set up a more general projectile Hamiltonian, 
including active and spectator electrons and their mutual Coulomb interactions. Treating 
these intra-molecular Coulomb correlations explicitly is however extremely demanding. It embraces
a full quantum-chemical description of the approaching molecule which cannot easily be adapted from 
one projectile to another. Since we are primarily interested in developing models and tools to 
be used for the description of secondary electron emission from plasma walls easy adaptation  
from one  projectile-target combination to another is however an important criterion for us. We stay 
therefore within the limits of a semi-empirical two-level system and use instead projection operators 
and two auxiliary bosons to assign and control the level energies. With these 
constructs it is possible to formulate a Hamiltonian containing both channels - \eqref{eq react rct} 
and~\eqref{eq react auger} - without introducing intra-molecular interactions. The projection 
operators allow us to assign different parameterizations to the two levels, depending on the occupancy 
of the system, whereas the auxiliary bosons enable us to mimic the intra-molecular Coulomb correlations 
which need to kick in to make the two tunneling processes involved in \eqref{eq react rct} resonant. 
Expressing the projection operators in terms of pseudo-particle operators with boson or fermion 
statistics opens then the door for employing non-equilibrium Green functions to derive from the Hamiltonian 
quantum-kinetic equations for the probabilities with which the molecular states can be found in the 
course of the collision.

The strength and flexibility of the pseudo-particle or slave field approach, originally developed by 
Coleman~\cite{Coleman1984New} in the context of the infinite-$U$ Anderson Hamiltonian, has been demonstrated 
many times for Anderson-type and Anderson-Newns-type 
models~\cite{Wingreen1994Anderson,Langreth1991Derivation,Shao1994Manybody,Aguado2003Kondo,Dutta2001Simple}. 
We apply this method to a generalized Anderson-Newns model describing the coupling of different molecular
configurations to a solid. It is of the type but not identical to the model introduced by 
Marston and coworkers~\cite{Onufriev1996Memory,Marston1993Manybody}. 
The derivation of the quantum-kinetic equations for the pseudo-particle propagators with the subsequent 
reduction to the rate equations for the occupancies of the molecular pseudo-particle states follows the 
work of Langreth and coworkers~\cite{Langreth1991Derivation,Shao1994Manybody}. As 
a result we obtain rate equations describing the de-excitation of \NitrogenDominantMetastableState\
in situations where the RCT and the Auger channel are simultaneously open. 
In the absence of the Auger channel the rate equations reduce 
to the ones we derived intuitively before for the isolated RCT channel~\cite{Marbach2012Resonant}.
Applying the model to the particular case of a diamond surface shows the resonant process dominating 
the Auger process. The overall secondary electron emission coefficient due to de-excitation of 
\NitrogenDominantMetastableState\ at a diamond surface is on the order of $10^{-1}$.

The outline of the rest of the paper is as follows. In Sec.~\ref{sec model} we describe the semi-empirical 
model on which our investigation of the de-excitation process is based. Thereafter, we explain in 
Sec.~\ref{sec slave boson} the pseudo-particle representation. Afterwards, we conduct in Sec.~\ref{sec quant kin} 
a second order quantum-kinetic calculation on top of the pseudo-particle model. In Sec.~\ref{sec semiclassical} 
we introduce a physically motivated semi-classical approximation that allows us to reduce the set of 
Dyson equations to a set of rate equations. Finally, we present in Sec.~\ref{sec results} the results 
for the diamond surface and conclude in Sec.~\ref{sec conclusion} with a brief summary of the main points of 
the work. Appendix~\ref{app langreth} lists the Langreth-Wilkins rules~\cite{Langreth1972Theory} as used in our 
calculation and Appendix~\ref{DysonEq} collects the second order Dyson equations for the molecular Green 
functions. 

\section{Model\label{sec model}}

The interacting molecule-surface system is characterized by three different types of electronic states: 
bound and unbound molecular states and states within the solid surface. In the spirit of our 
previous work~\cite{Marbach2011Auger,Marbach2012Resonant} we restrict the attention to those states 
whose occupancies change during the molecule-surface collision. For these states and the coupling 
between them we construct a semi-empirical model. Its matrix elements can be either obtained from 
quantum-mechanical calculations based on particular assumptions about the electron wave functions 
and/or experimentally measured ionization energies, electron affinities, surface response functions, 
and electron tunneling rates. Since we are primarily interested in the quantum-kinetic handling of 
the semi-empirical model we pursue for simplicity the former route. A more realistic parameterization 
of the model is however in principle possible.
\begin{figure}
  \centerline{\includegraphics[scale=1]{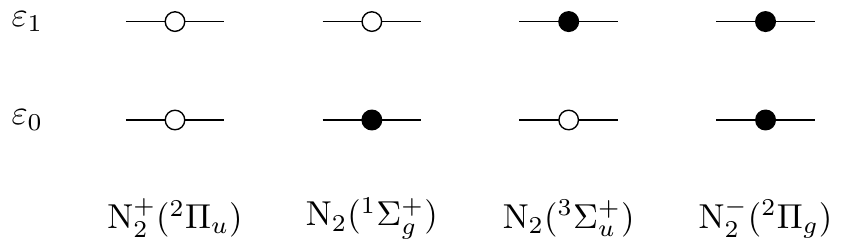}}
  \caption{Correspondence between the occupation of the two level system and the molecular states.
  Here~$\eps_0$ and~$\eps_1$ denote the energies of the levels~$0$ and $1$, respectively.}
  \label{fig two level system}
\end{figure}

We treat the relevant bound states of the nitrogen molecule in terms of a two-level system consisting 
of a ground state level~"$0$" and an excited level~"$1$". Within a linear combination of atomic orbital 
(LCAO) description of the molecule these two levels represent the nitrogen molecule's $2\pi_u$ and~$2\pi_g$ 
orbitals. Each of the two orbitals can carry four electrons. We neglect however the three electrons 
in the $2\pi_u$ orbital and the three holes in the $2\pi_g$ orbital which are not directly involved 
in the de-excitation process. They act only as frozen-in spectators. For the same reason we neglect 
the electron spin and treat the magnetic quantum number~${m=\pm1}$ as an initial parameter. Hence, both 
levels of our model carry at most one electron.

\begin{figure}
  \centerline{\includegraphics[scale=1]{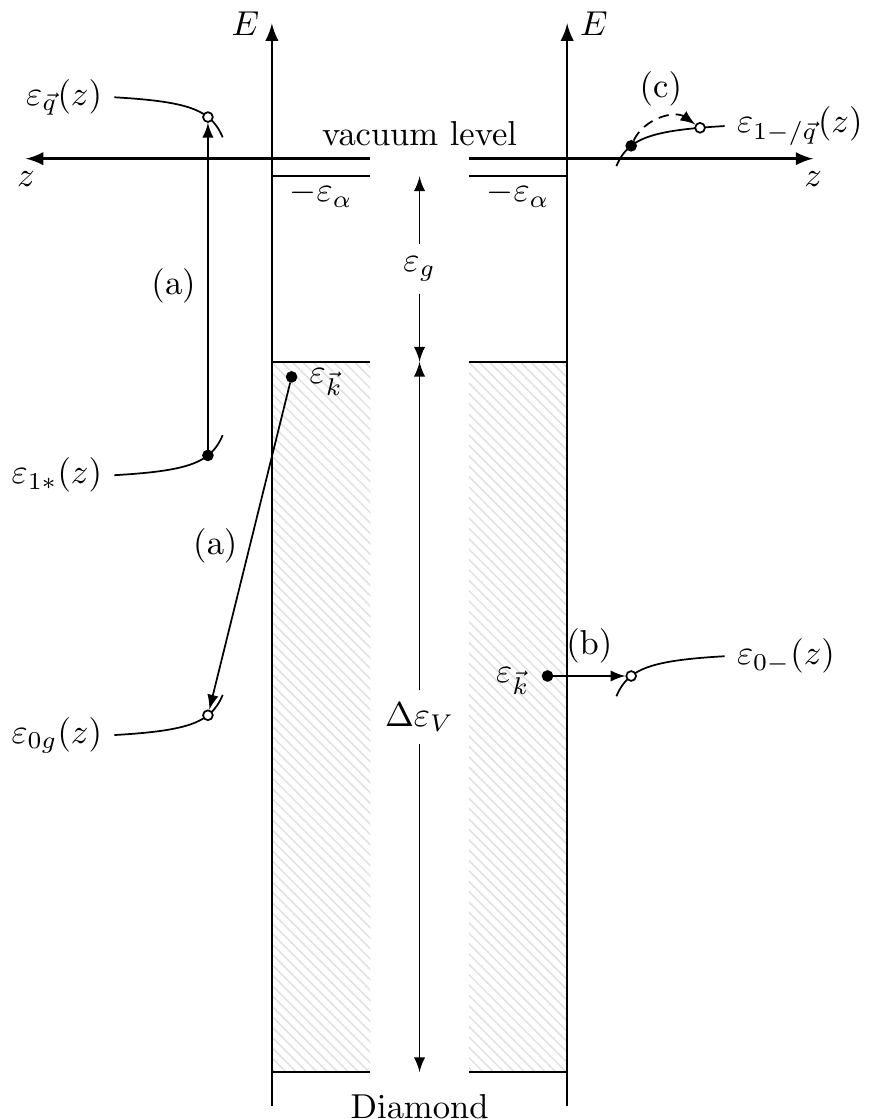}}
  \caption{Energy scheme for the case of a metastable \NitrogenDominantMetastableState\ molecule interaction
with a diamond surface. The Auger de-excitation channel (a) is depicted on the left hand side whereas
the RCT electron capture (b) and the subsequent RCT electron release (c) are shown on the right hand side.
The latter transition is drawn with a dashed line because the electron emission actually occurs at the same $z$~position
and hence energy. It is resonant. Therefore the emitted electron will only reach the shown location over time.
The drawing is to scale in terms of energy units. \label{fig energy scheme}}
\end{figure}

The two-level system represents any of the molecular states depicted in Fig.~\ref{fig two level system}.
The positive ion \NitrogenPositiveIon, the ground state \NitrogenGroundState, the metastable 
state \NitrogenDominantMetastableState, and the negative ion \NitrogenNegativeIonResonance. 
Depending on the particular occupation the energies $\eps_0$ and~$\eps_1$ correspond therefore to the 
ionization energies of different molecular states. Due to intra-molecular Coulomb interactions
these ionization energies are in general different~\cite{Kaldor1984Generalmodelspace,Honigmann2006Complex}. 
We have to allow therefore $\eps_0$ and~$\eps_1$ to depend on the
occupancy of the two-level system, that is, on the molecular state it is supposed to represent. In 
addition, the ionization energies are also subject to the surface's image potential $V_i(z)$ and 
thus vary with time as the molecule moves with respect to the surface. Using the analysis presented in 
Ref.~\onlinecite{Marbach2012Resonant} we find
\begin{subequations}
\begin{align}
  \eps_{0g}(z) & = \eps_{0g}^\infty - V_i(z) \;,\\
  \eps_{1*}(z) & = \eps_{1*}^\infty - V_i(z) \;,\\
  \eps_{0-}(z) & = \eps_{0-}^\infty + V_i(z) \;, \\
  \eps_{1-}(z) & = \eps_{1-}^\infty + V_i(z) \;,
\end{align}\label{eq energy shifts}%
\end{subequations}
where the subscripts~$g$, $*$ and $-$ signal the dependence of the energy levels on the molecular 
state denoting, respectively, the ground state molecule, the metastable molecule, and the negative 
ion. The unperturbed molecular energies are given by~\cite{Kaldor1984Generalmodelspace,Honigmann2006Complex}
\begin{equation}
\begin{split}
  \eps_{0g}^\infty & = -17.25\, eV\;, \phantom{\eps_{1-}^\infty} \eps_{1*}^\infty = -9.67\, eV \;,\\
  \eps_{0-}^\infty & = -14.49\, eV\;, \phantom{\eps_{1*}^\infty} \eps_{1-}^\infty = 1.18\, eV\;.
\end{split}\label{eq inf mol energies}%
\end{equation}
The overall energy scheme of the coupled molecule-surface system is sketched in Fig.~\ref{fig energy scheme}
for the particular case of a diamond surface. As can be seen the positive ion is neither involved in the 
RCT nor the Auger process. 
%The positive ion's ionization energies are therefore not listed in
%\eqref{eq energy shifts}.
%

For the image potential we employ for simplicity the classical expression,
\begin{equation}
  V_i(z) \approx - \frac{\eps_r^b - 1}{\eps_r^b + 1} \frac{e^2}{16 \pi \eps_0} \frac{1}{z} \;,
\end{equation}
with $\eps_r^b$ standing for the surface's static bulk dielectric constant. Close to the surface the 
image potential is however truncated according to $V_i(z_c)=V_0$ where $V_0$ is the depth of the 
potential barrier confining the electrons of the solid participating in the de-excitation process. As 
in our previous investigations~\cite{Marbach2011Auger,Marbach2012Resonant}, we describe the solid by 
a step potential of depth $V_0$. For a metallic surface~\cite{Marbach2011Auger} the step depth is the 
width of the conduction band $\Delta\eps_C$, that is, the sum of the work function $\Phi_W$ and the Fermi 
energy  $\eps_F$ whereas for a dielectric surface~\cite{Marbach2012Resonant} it is the sum of the width 
of the valence band $\Delta\eps_V$, the energy gap $\eps_g$, and the electron affinity~$\eps_\alpha$ 
which can be positive or negative. 

In both the RCT and the Auger channel the emitted electron stems from the molecule and, thus, the emission proceeds 
into the molecular continuum states. We model the latter as free electron states moving along with the molecule and 
label them with $\vec{q}$. Electrons residing in those states are also affected by the image potential and their 
energy is thus given by
\begin{equation}
  \eps_{\vec{q}}(z) = \eps_{\vec{q}}^\infty + V_i(z) = \frac{\hbar^2 q^2}{2 m_e} + V_i(z) \;.
  \label{eq eps q}
\end{equation}
The wave function of the emitted electron is a two-center Coulomb wave for the Auger channel~\cite{Marbach2011Auger} 
and a plane wave for the RCT channel~\cite{Marbach2012Resonant}. The latter is a special case of the former which
holds for zero effective nucleus charge. Since in the RCT channel the emitted electron leaves a neutral molecule
behind the plane wave is the suitable choice for this channel. In the Auger channel however the emitted electron 
feels the residual two-center Coulomb attraction of the ion core. The two-center Coulomb wave takes this effect 
into account. To complete the description of the model we note that we use LCAO molecular wave functions for 
the two-level system and eigenstates of a step potential for the single-electron wave functions of the solid.
Explicit expressions for the wave functions and the details of the calculation of the Auger and tunneling matrix 
elements are given in Refs.~\onlinecite{Marbach2011Auger,Marbach2012Resonant}.

Clearly, the model just described is based on a simple surface potential. Modeling the surface potential 
as a superposition of a step potential (which somewhat underestimates the exponential tail of the metal 
electron's wave functions and hence the absolute value of the matrix elements) and a classical image potential 
(which is less critical because the turning point turns out to be far in front of the surface) allows us 
however to write down analytic expressions for the matrix elements. For the quantum kinetics itself the 
matrix elements are only parameters. Our approach can thus be also furnished with improved matrix elements 
obtained from more realistic potentials. Based on the work of \citeauthor{Kurpick1996Basic}~\cite{Kurpick1996Basic} 
we would expect however not too dramatic differences.

We now cast the semi-empirical model just described into a mathematical form. Introducing projection operators 
\begin{equation}
P^{n_0n_1}=|n_0n_1\rangle\langle n_0n_1|
\label{projectors}
\end{equation}
projecting onto states of the two-level system with $n_0=0,1$ electrons in the lower and $n_1=0,1$ electrons in 
the upper state, the transitions shown in Fig.~\ref{fig energy scheme} give rise to a generalized Anderson-Newns 
model~\cite{Marston1993Manybody},
\begin{equation}
\begin{split}
  H(t) & = \sum_{\vec k} \eps_{\vec k}^{\phdag} \, c_{\vec{k}}^{\dag} \, c_{\vec{k}}^{\phdag} 
       + \sum_{\vec q} \eps_{\vec q}^{\phdag}(t) \, c_{\vec{q}}^{\dag} \, c_{\vec{q}}^{\phdag} \\
     & + \omega_0 b_0^{\dag} b_0 + \omega_1 b_1^{\dag}b_1\\
     &  + \sum_{n_0,n_1} P^{n_0n_1} \left(\eps^{n_0n_1}_{0}(t) \, c_{0}^{\dag} \, c_{0}^{\phdag} 
        + \eps^{n_0n_1}_{1}(t) \, c_{1}^{\dag} \, c_{1}^{\phdag} \right) \\
     &  + \sum_{\vec k} \left(\big[P^{01}+P^{11}\big]\RCTCaptureME^{\phconj\!}(t) \, c_{\vec{k}}^{\dag}\, b_0^{\dag}\, c_{0}^{\phdag} 
       + h.c. \right) \\
     & + \sum_{\vec q} \left(\big[P^{10}+P^{11}\big]\RCTSIReleaseME^{\phconj\!}(t) \, c_{\vec{q}}^{\dag}\, b_1^{\dag}\, c_{1}^{\phdag} 
       + h.c. \right) \\
  & + \sum_{\vec{k} \vec{q}} \left(\big[P^{10}+P^{01}\big] \DPDME(t) \, 
  c_0^\dag \, c_{\vec{k}}^\phdag \, c_{\vec{q}}^\dag \, c_1^\phdag + h.c. \right) \;,
\end{split}\label{eq hamiltonian}%
\end{equation}
where  $\eps^{10}_{0}=\eps_{0g}$, $\eps^{01}_{1}=\eps_{1*}$, $\eps^{11}_{0}=\eps_{0-}$, and
$\eps^{11}_{1}=\eps_{1-}$. The remaining energy levels need not be specified further. They drop 
out in the course of the pseudo-particle representation presented in the next section. The two auxiliary 
bosons $b_0^{(\dag)}$ and $b_1^{(\dag)}$ mimic 
the intra-molecular Coulomb correlations which make the two steps of the RCT channel \eqref{eq react rct} 
resonant. This can be accomplished by setting $\omega_0=\eps_1^{11}-\eps_1^{01}$ and $\omega_1=\eps_0^{11}-\eps_0^{10}$. 
The initial energy of the tunneling electron is then resonant
respectively with the lower and the upper level of the negative ion. The rest of the Hamiltonian is 
written in the notation we used in our previous work~\cite{Marbach2011Auger,Marbach2012Resonant}. 

The time dependence of the Hamiltonian arises from the trajectory of the molecule's center of mass~${\vec{R}(t)}$. 
For simplicity we assume normal incidence described by the trajectory
$\vec{R}(t) = \bigl( v_0 \abs{t} + z_0 \bigr) \vec{e}_z$,
where~$v_0$ is a constant velocity and~$z_0$ is the molecule's turning point which can be calculated from a Morse type 
interaction potential~\cite{Marbach2011Auger}. Employing the trajectory the~$z$~dependence 
of the molecular energies~\eqref{eq energy shifts} and the energy of the emitted electron~\eqref{eq eps q} 
transforms into a time dependence. Similarly, the Auger matrix element $\DPDME$ and the two resonant tunneling 
matrix elements $\RCTCaptureME$ and $\RCTSIReleaseME$ acquire also a time-dependence. 

The quantum-kinetic calculation presented below treats the matrix elements of the Hamiltonian as 
parameters. We are thus not restricted to the specific approximations to the matrix elements derived in 
our previous work~\cite{Marbach2011Auger,Marbach2012Resonant}. We could as well use matrix elements
from ab-initio calculations or experimental measurements. The natural decay of the negative ion, 
described by the rate $\Gamma_n=1/\tau_n$ with $\tau_n$ the natural lifetime of the 
negative ion, is not included in the Hamiltonian~\eqref{eq hamiltonian}. It will be inserted at the end 
into the final set of rate equations for the molecular occupancies.

\section{Pseudo-particle representation\label{sec slave boson}}

The projection operators \eqref{projectors} permit us to describe the  
transitions~\eqref{eq react rct} and~\eqref{eq react auger} by a single Hamiltonian. Depending on the 
process and thus the occupancy of the molecular levels different matrix elements can be assigned to the 
Hamiltonian. The projectors also guarantee that the~\NitrogenPositiveIon\ state never occurs. In 
other words, they ensure that the occupancies of the two molecular levels never vanishes simultaneously. 
For instance, an electron residing in the upper level can only be released when an electron has been 
captured in the lower level. 

A drawback of the projection operators is that they are not suitable for a diagrammatic treatment which
on the other hand is a powerful tool to set up quantum-kinetic equations. To
remedy this drawback we now employ a pseudo-particle approach to express the Hamiltonian \eqref{eq hamiltonian} 
in terms of slave 
fields~\cite{Coleman1984New,Wingreen1994Anderson,Langreth1991Derivation,Shao1994Manybody,Aguado2003Kondo,Dutta2001Simple}. 
The starting point for this procedure is the completeness condition, 
\begin{equation}
  |0 0 \rangle \langle 0 0| + |1 0 \rangle \langle 1 0| + |0 1 \rangle \langle 0 1| + |1 1 \rangle \langle 1 1| = 1\;,
  \label{eq-completeness}
\end{equation}
which expresses the fact that the molecule can be only in either one of the configurations depicted in
Fig.~\ref{fig two level system}. Introducing pseudo-particle operators $c_+^\dag$, $c_g^\dag$, $c_*^\dag$,
and $c_-^\dag$ that create the positive ion, the ground state molecule, the metastable molecule, and the 
negative ion from an abstract vacuum state,
\begin{subequations}
\begin{align}
  |0 0 \rangle & = c_+^\dag | \text{vac} \rangle, \; |1 0 \rangle = c_g^{\dag\,\,} | \text{vac} \rangle \;, \\
  |0 1 \rangle & = c_*^{\dag\,\,} | \text{vac} \rangle, \; |1 1 \rangle = c_-^\dag | \text{vac} \rangle \;, 
\end{align}\label{eq-basis-states}%
\end{subequations}
the completeness condition~\eqref{eq-completeness} becomes 
\begin{equation}
  c_+^\dag \, c_+^\phdag + c_g^\dag \, c_g^\phdag + c_*^\dag \, c_*^\phdag + c_-^\dag \, c_-^\phdag = 1 \;.
  \label{eq-constraint-full}
\end{equation}
Using~\eqref{eq-completeness} and~\eqref{eq-basis-states} the operators~$c_{0/1}^{(\dag)}$ creating and destroying 
an electron in the two states of the two-level system can then be written as
\begin{subequations}
\begin{align}
  c_0^\phdag & = c_0^\phdag * 1 = |0 0 \rangle \langle 1 0| - |0 1 \rangle \langle 1 1| = c_+^\dag \, c_g^\phdag - c_*^\dag \, c_-^\phdag \;, \\
  c_0^\dag & = c_0^\dag * 1 = |1 0 \rangle \langle 0 0| - |1 1 \rangle \langle 0 1| = c_g^\dag \, c_+^\phdag - c_-^\dag \, c_*^\phdag \;, \\
  c_1^\phdag & = c_1^\phdag * 1 = |0 0 \rangle \langle 0 1| + |1 0 \rangle \langle 1 1| = c_+^\dag \, c_*^\phdag + c_g^\dag \, c_-^\phdag \;, \\
  c_1^\dag & = c_1^\dag * 1 = |0 1 \rangle \langle 0 0| + |1 1 \rangle \langle 1 0| = c_*^\dag \, c_+^\phdag + c_-^\dag \, c_g^\phdag \;.
\end{align}\label{eq-c01-decomposition}%
\end{subequations}
In order to satisfy the anti-commutation relations of the~$c_{0/1}^{(\dag)}$ one has to define~${c_0^\phdag \, 
|1 1 \rangle = - |0 1 \rangle}$ and~${c_0^\dag \, |0 1 \rangle = - |1 1 \rangle}$ (see also 
Ref.~\onlinecite{Carmona2002Slave}). The anti-commutator relations then reproduce the completeness 
condition~\eqref{eq-constraint-full} when either the~$c_{g/*}$ are bosonic and the~$c_{-/+}$ are fermionic or 
the~$c_{g/*}$ are fermionic and the~$c_{-/+}$ are bosonic. Without loss of generality we choose $c_g$ and $c_*$ 
to be bosonic and declare the labeling conventions
\begin{subequations}
\begin{align}
  c_g^{(\dag)} \longrightarrow b_g^{(\dag)} \;, \phantom{c_+^{(\dag)} \longrightarrow f_+^{(\dag)}} & \hspace{-1.8cm}\phantom{c_-^{(\dag)}} c_*^{(\dag)} \longrightarrow b_*^{(\dag)} \;,\\
  c_+^{(\dag)} \longrightarrow f_+^{(\dag)} \;, \phantom{c_g^{(\dag)} \longrightarrow b_g^{(\dag)}} & \hspace{-1.8cm}\phantom{c_*^{(\dag)}} c_-^{(\dag)} \longrightarrow f_-^{(\dag)} \;.
\end{align}
\label{convention}
\end{subequations}
The constraint~\eqref{eq-constraint-full} reduces then to
\begin{equation}
  Q = b_g^\dag \, b_g^\phdag + b_*^\dag \, b_*^\phdag + f_-^\dag \, f_-^\phdag + f_+^\dag \, f_+^\phdag = 1 \;,
  \label{eq-projected-constraint}
\end{equation}
where we have introduced the usual pseudo-particle number operator~$Q$ to encapsulate the constraint. 

Formally, the auxiliary fermion and boson operators are pseudo-particle operators creating and annihilating
molecular configurations. The constraint ensures that at any time only one of the four possible molecular 
configurations is present in the system. The occupancy of a molecular pseudo-particle 
state is thus at most unity. Hence, it represents the probability with which the molecular 
configuration it describes appears in the course of the scattering event. 

Inserting the decomposition~\eqref{eq-c01-decomposition} into the Hamiltonian \eqref{eq hamiltonian}, 
making the identifications \eqref{convention}, and collecting only terms which are
in accordance with \eqref{eq-projected-constraint} is straightforward. The result is 
\begin{equation}
\begin{split}
  H(t) = & \sum_{\vec k} \eps_{\vec k}^{\phdag} \, c_{\vec{k}}^{\dag} \, c_{\vec{k}}^{\phdag} + \sum_{\vec q} \eps_{\vec q}^{\phdag}(t) \, c_{\vec{q}}^{\dag} \, c_{\vec{q}}^{\phdag} \\
  & + \omega_0 b_0^{\dag} b_0 + \omega_1 b_1^{\dag}b_1\\
  & + \eps_g^{\phdag}(t) \, b_{g}^{\dag} \, b_{g}^{\phdag} + \eps_*^{\phdag}(t) \, b_{*}^{\dag} \, b_{*}^{\phdag} + \eps_-^{\phdag}(t) \, f_{-}^{\dag} \, f_{-}^{\phdag} \\[1.5ex]
  & - \sum_{\vec k} \Bigl( \RCTCaptureME^{\phconj\!}(t) \, c_{\vec{k}}^{\dag} \, b_0^{\dag}\, b_{*}^{\dag} \, f_{-}^{\phdag} + h.c. \Bigr) \\
  & + \sum_{\vec q} \Bigl( \RCTSIReleaseME^{\phconj\!}(t) \, c_{\vec{q}}^{\dag} \, b_1^{\dag}\, b_{g}^{\dag} \, f_{-}^{\phdag} + h.c. \Bigr) \\
  & + \sum_{\vec{k} \vec{q}} \left( \DPDME(t) \, c_{\vec{k}}^\phdag \, c_{\vec{q}}^\dag \, b_g^\dag \, b_*^\phdag + h.c. \right) \;,
\end{split}\label{eq-projected-hamiltonian}%
\end{equation}
where we introduced the abbreviations
\begin{equation}
  \eps_g = \eps_0^{10} \;,\quad \eps_* = \eps_1^{01} \;,\quad \eps_- = \eps_0^{11} + \eps_1^{11} \;.
\label{assigment}
\end{equation}
Notice, no term in the Hamiltonian contains the operator $f_+$ or its adjoint. This is by construction
because the positive ion \NitrogenPositiveIon\ is not involved in the two transitions the Hamiltonian 
is supposed to model. The physical meaning of the various terms in the Hamiltonian is particularly 
transparent. Consider for instance the last term describing the Auger de-excitation. A metastable 
molecule and an electron from the surface disappear while a ground state molecule and an Auger 
electron are created. 

The operators $f_-$ and $b_{*/g}$ comply to standard commutation and anti-commutation relations. 
It is thus possible to conduct a non-equilibrium diagrammatic expansion of the interaction terms 
in~\eqref{eq-projected-hamiltonian}. The Hamiltonian~\eqref{eq-projected-hamiltonian} conserves 
the pseudo-particle number~$Q$. The quantum-kinetic equations however may contain terms which violate 
the constraint. The projection onto the physical subspace with~${Q=1}$ needs to be therefore carried out 
explicitly. For this purpose we employ in the next section the Langreth-Nordlander projection 
technique~\cite{Langreth1991Derivation}.

\section{Quantum-kinetics\label{sec quant kin}}

To start the quantum-kinetic calculation we first define a contour-ordered fermion Green function $G_-$ 
for the negative ion and contour-ordered boson Green functions $B_{*/g/0/1}$ for the metastable molecule,
the molecular ground state, and the two auxiliary bosons, respectively. Using the notation 
of \citeauthor{Langreth1991Derivation}~\cite{Langreth1991Derivation} we write
\begin{subequations}
\begin{align}
  i G_-(t,t^\prime) & = \bigl\langle T_\mathcal{C} \, f_-^\phdag(t) \, f_-^\dag(t^\prime) \bigr\rangle \;, \\
  i B_{l}(t,t^\prime) & = \bigl\langle T_\mathcal{C} \, b_{l}^\phdag(t) \, b_{l}^\dag(t^\prime) \bigr\rangle \;,
\end{align}
\label{contourGF}
\end{subequations}
where $l=*,g,0,1$, and define the analytic pieces~$G_-^{\genfrac{}{}{0pt}{}{>}{<}}$ and 
$B_{l}^{\genfrac{}{}{0pt}{}{>}{<}}$ by  
\begin{subequations}
\begin{align}
\begin{split}
  i G_-(t,t^\prime) & = \Theta_\mathcal{C}(t-t^\prime) \, G_-^>(t,t^\prime) \\
  & \phantom{=} - \Theta_\mathcal{C}(t^\prime-t) \, G_-^<(t,t^\prime) \;,
\end{split} \\
\begin{split}
  i B_{l}(t,t^\prime) & = \Theta_\mathcal{C}(t-t^\prime) \, B_{l}^>(t,t^\prime) \\
  & \phantom{=} + \Theta_\mathcal{C}(t^\prime-t) \, B_{l}^<(t,t^\prime) \;.
\end{split}
\end{align}%
\end{subequations}
The time-ordering operator~$T_\mathcal{C}$ and the~$\Theta_\mathcal{C}$~function are defined on a complex time contour. 
The associated retarded Green functions~$G_-^R$ and~$B_{l}^R$ are given by
\begin{subequations}
\begin{align}
  i G_-^R(t,t^\prime) & = \Theta(t-t^\prime) \bigl( G_-^>(t,t^\prime) + G_-^<(t,t^\prime) \bigr) \;, \\
  i B_{l}^R(t,t^\prime) & = \Theta(t-t^\prime) \bigl( B_{l}^>(t,t^\prime) - B_{l}^<(t,t^\prime) \bigr)\;,
\end{align}\label{eq-retarded-definition}%
\end{subequations}
whereas the advanced functions~$G_-^A$ and~$B_{l}^A$ may be constructed from~\eqref{eq-retarded-definition}  
using the relations
\begin{subequations}
\begin{align}
  G_-^A(t,t^\prime) & = \bigl[ G_-^R(t^\prime,t) \bigr]^\conj \;, \\
  B_{l}^A(t,t^\prime) & = \bigl[ B_{l}^R(t^\prime,t) \bigr]^\conj \;.
\end{align}
\label{SymmetryRelations}
\end{subequations}
In contrast to our previous work~\cite{Marbach2011Auger,Marbach2012Resonant}, we no longer use Keldysh's 
matrix notation. Langreth's notation is more convenient. It enables us to use the powerful Langreth-Wilkins
rules~\cite{Langreth1972Theory} which lead more directly to the quantum-kinetic equations for the 
molecular occupancies, that is, the probabilities with which the molecular configurations involved in the
de-excitation process appear in the course of the scattering event.

\begin{figure}
\centerline{\includegraphics[scale=1]{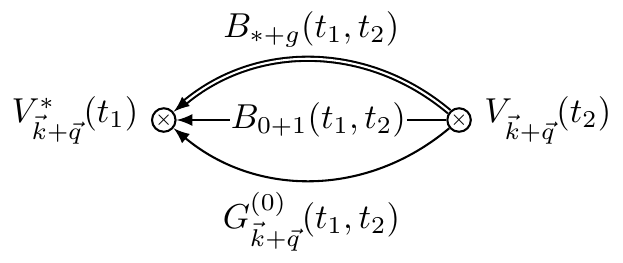}}
\centerline{\includegraphics{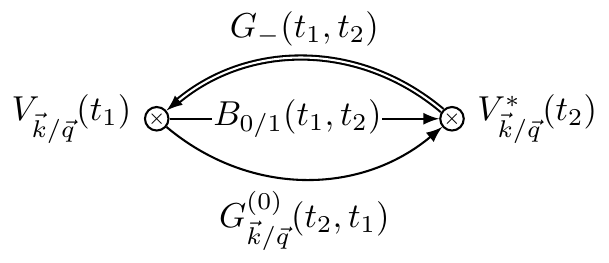}}
\centerline{\includegraphics{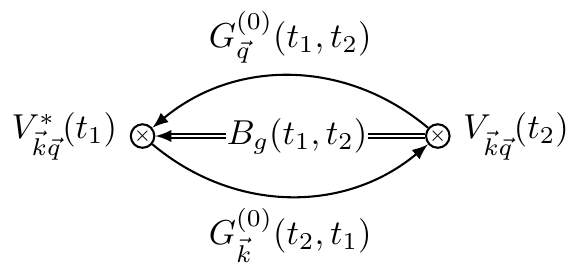}}
\centerline{\includegraphics{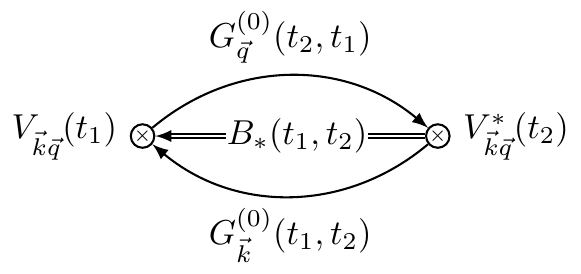}}
\caption{Second order self-energy diagrams in self-consistent non-crossing approximation. Depicted are from top to 
bottom the self-energy of the fermionic level~$-i\Sigma_-(t_1,t_2)$, the RCT component of the self-energies of the 
bosonic levels~$-i\Pi_{*}(t_1,t_2)$ and~$-i\Pi_{g}(t_1,t_2)$, the Auger component of the metastable boson
self-energy~$-i\Pi_*(t_1,t_2)$, and the Auger component of the ground state boson self-energy~$-i\Pi_g(t_1,t_2)$.
In the upper most diagram the $+$ sign in the indices means that there are two separate diagrams, one with
the~$\vec{k}$ and~$*$~indices and one with the~$\vec{q}$ and~$g$~indices, both of which contribute in an additive way
and have thus to be summed. Furthermore in the diagram below the upper most one the subscripts~${\vec{k}}$ 
and $0$ hold for~$\Pi_*$ and the subscripts~${\vec{q}}$ and $1$ hold for~$\Pi_g$. In all diagrams a double line 
indicates a full propagator whereas a single line denotes an unperturbed propagator.
\label{fig self energies}}
\end{figure}

In order to calculate the self-energies we truncate the diagrammatic expansion beyond the second order and 
employ the self-consistent non-crossing approximation~\cite{Langreth1991Derivation}.
The diagrams for the fermionic self-energy~$\Sigma_-$ and the 
bosonic self-energies~$\Pi_{*}$ and~$\Pi_{g}$ are shown in Fig.~\ref{fig self energies}. 
Mathematically they read
\begin{subequations}
\begin{align}
\begin{split}
  \Sigma_-(t_1,t_2) & = i\, \sigma_{\vec{k}}(t_1,t_2) \, B_*(t_1,t_2) \\
  & \phantom{=} \,\, +i \, \sigma_{\vec{q}}(t_1,t_2) \, B_g(t_1,t_2) \;,
\end{split} \\
\begin{split}
  \Pi_{*}(t_1,t_2) & = -i\, \sigma_{\vec{k}}(t_2,t_1) \, G_-(t_1,t_2) \\
  & \phantom{=} \,\, + \sigma_{\vec{k}\vec{q}}(t_1,t_2) \, B_g(t_1,t_2) \;,
\end{split} \\
\begin{split}
  \Pi_{g}(t_1,t_2) & = -i\, \sigma_{\vec{q}}(t_2,t_1) \, G_-(t_1,t_2) \\
  & \phantom{=} \,\, + \sigma_{\vec{k}\vec{q}}(t_2,t_1) \, B_*(t_1,t_2) \;
\end{split}
\end{align}\label{eq unprojected self energies}
\end{subequations}
with 
\begin{subequations}
\begin {align}
\begin{split}
  \sigma_{\vec{k}/\vec{q}}(t_1,t_2) & = \frac{i}{\hbar^2} \sum_{\vec{k}/\vec{q}} V_{\vec{k}/\vec{q}}^\conj(t_1) \, V_{\vec{k}/\vec{q}}^\phconj(t_2) \\
  & \phantom{=} \times G_{\vec{k}/\vec{q}}^{(0)}(t_1,t_2) B^{(0)}_{0/1}(t_1,t_2)\;,
\end{split} \\
\begin{split}
  \sigma_{\vec{k}\vec{q}}(t_1,t_2) & = - \frac{1}{\hbar^2} \sum_{\vec{k}\vec{q}} V_{\vec{k}\vec{q}}^\conj(t_1) \, V_{\vec{k}\vec{q}}^\phconj(t_2) \\
  & \phantom{=} \times G_{\vec{k}}^{(0)}(t_2,t_1) \, G_{\vec{q}}^{(0)}(t_1,t_2) \;,
\end{split}
\end{align}\label{eq sigma definitions}%
\end{subequations}
and $G_{\vec{q}/\vec{k}}^{(0)}$ and $B_{0/1}^{(0)}$ denoting, respectively, the contour-ordered Green functions 
for a free/valence band electron and an auxiliary boson. 

In the self-energies~\eqref{eq unprojected self energies} the two reaction channels~\eqref{eq react rct} 
and~\eqref{eq react auger} are separated. Every term involving~$\sigma_{\vec{k}}$ or~$\sigma_{\vec{q}}$ 
refers to the RCT channel and every term containing~$\sigma_{\vec{k}\vec{q}}$ pertains to the Auger channel. 
Due to the dressed Green functions the two channels are however coupled. The Green functions of the
auxiliary bosons contained in~$\sigma_{\vec{k}}$ and $\sigma_{\vec{q}}$ ensure that the two tunneling
processes contained in~\eqref{eq react rct} are resonant. Physically, the auxiliary bosons simulate 
the action of intra-molecular correlations which kick-in when an electron hops to-and-fro the molecule. 

Using the Langreth-Wilkins rules for analytic continuation (see Refs.~\onlinecite{Langreth1972Theory,Langreth1991Derivation} 
and Appendix~\ref{app langreth}) we obtain from the self-energies~\eqref{eq unprojected self energies} the set of Dyson 
equations given in Appendix~\ref{DysonEq}. The components of these equations arising from the RCT terms are equivalent to 
the ones in Ref.~\onlinecite{Langreth1991Derivation} but with two bosonic pseudo-particles instead of one and an energy 
shift caused by the auxiliary bosons. The set of Dyson 
equations~\eqref{eq-plain-dyson} contains terms which violate the constraint~\eqref{eq-projected-constraint}. Before  
physically meaningful information can be extracted the Dyson equations have to be projected onto the physical subspace 
defined by the constraint~\eqref{eq-projected-constraint}. 

The procedure to achieve this is originally due to Langreth and Nordlander and has been outlined several 
times~\cite{Langreth1991Derivation,Wingreen1994Anderson,Aguado2003Kondo}.
It is based on inspecting the order of the Green functions in the conserved pseudo-particle number $Q$. The retarded 
functions $G_-^R$ and $B_{*/g}^R$ are proportional to $Q^0$ while the lesser Green functions~$G_-^<$ and~$B_{*/g}^<$ are 
proportional to~$Q^1$. Thus, we have to omit any terms of higher order than~$Q^0$ from the retarded self-energies and any 
terms of higher order than~$Q^1$ from the lesser self-energies. This approach is not an additional approximation but an 
exact projection enforced by the constraint~\cite{Langreth1991Derivation}. 

Before carrying out the projection we split off the Green functions' oscillating factors by means of the 
decompositions~\cite{Shao1994Manybody}
\begin{subequations}
\begin{align}
  G_-^{</R/A}(t,t^\prime) & = \widetilde{G}_-^{</R/A}(t,t^\prime) \, e^{-\frac{i}{\hbar} \int_{t^\prime}^t \mathrm{d}t_1 \, \eps_-(t_1)} \;, \\
  B_*^{</R/A}(t,t^\prime) & = \widetilde{B}_*^{</R/A}(t,t^\prime) \, e^{-\frac{i}{\hbar} \int_{t^\prime}^t \mathrm{d}t_1 \, \eps_*(t_1)} \;, \\
  B_g^{</R/A}(t,t^\prime) & = \widetilde{B}_g^{</R/A}(t,t^\prime) \, e^{-\frac{i}{\hbar} \int_{t^\prime}^t \mathrm{d}t_1 \, \eps_g(t_1)} \;,
\end{align}\label{eq-g<-decomposition}%
\end{subequations}
and
\begin{subequations}
\begin{align}
  \widetilde{G}_-^R(t,t^\prime) & = -i \, \Theta(t-t^\prime) \, g_-(t,t^\prime) \;, \\
  \widetilde{G}_-^A(t,t^\prime) & = i \, \Theta(t^\prime - t) \, g_-(t,t^\prime) \;, \\
  \widetilde{B}_{*/g}^R(t,t^\prime) & = -i \, \Theta(t-t^\prime) \, b_{*/g}(t,t^\prime) \;, \\
  \widetilde{B}_{*/g}^A(t,t^\prime) & = i \, \Theta(t^\prime - t) \, b_{*/g}(t,t^\prime) \;.
\end{align}\label{eq-gr-decomposition}%
\end{subequations}
Using the definition of the retarded and advanced Green functions we find the following relations~\cite{Shao1994Manybody}
\begin{subequations}
\begin{align}
  g_-(t,t) & = b_{*/g}(t,t) = 1 \;, \\
  g_-(t,t^\prime) & = \bigl[ g_-(t^\prime,t) \bigr]^\conj \;, \\
  b_{*/g}(t,t^\prime) & = \bigl[ b_{*/g}(t^\prime,t) \bigr]^\conj \;.
\end{align}
\end{subequations}
Within the Dyson equations the oscillating terms emerging from Eqs.~\eqref{eq-g<-decomposition} will be absorbed in the functions~$\widetilde{\sigma}_{\vec{k}}^<$, $\widetilde{\sigma}_{\vec{q}}^>$ and~$\widetilde{\sigma}_{\vec{k}\vec{q}}^>$ which are defined by
\begin{subequations}
\begin{align}
  \sigma_{\vec{k}}^{<}(t,t^\prime) & = \widetilde{\sigma}_{\vec{k}}^{<}(t,t^\prime) \, e^{-\frac{i}{\hbar} \int_{t^\prime}^t \mathrm{d}t_1 \, [ \eps_-(t_1) - \eps_*(t_1) ]} \;, \\
  \sigma_{\vec{q}}^{>}(t,t^\prime) & = \widetilde{\sigma}_{\vec{q}}^{>}(t,t^\prime) \, e^{-\frac{i}{\hbar} \int_{t^\prime}^t \mathrm{d}t_1 \, [ \eps_-(t_1) - \eps_g(t_1) ]} \;, \\
  \sigma_{\vec{k}\vec{q}}^{>}(t,t^\prime) & = \widetilde{\sigma}_{\vec{k}\vec{q}}^{>}(t,t^\prime) \, e^{-\frac{i}{\hbar} \int_{t^\prime}^t \mathrm{d}t_1 \, [ \eps_*(t_1) - \eps_g(t_1) ]} \;.
\end{align}\label{eq-sigma-decomposition}%
\end{subequations}
The terms~$\sigma_{\vec{k}}^>$, $\sigma_{\vec{q}}^<$, and~$\sigma_{\vec{k}\vec{q}}^<$ vanish identically due to the 
initial conditions $n_0(t_0)=n_{\vec{k}}(t_0)=1$ and $n_1(t_0)=n_{\vec{q}}(t_0)=0$ since
\begin{subequations}
\begin{align}
  \sigma_{\vec{k}}^>(t,t^\prime) & \sim \bigl( 1 - n_{\vec{k}}(t_0) \bigr) \bigl( 1 - n_0(t_0) \bigl) = 0 \;, \\
  \sigma_{\vec{q}}^<(t,t^\prime) & \sim n_{\vec{q}}(t_0) n_1(t_0) = 0 \;, \\
  \sigma_{\vec{k}\vec{q}}^<(t,t^\prime) & \sim n_{\vec{q}}(t_0) = 0 \;.
\end{align}\label{eq-sigma-initial}%
\end{subequations}
Employing the Langreth-Nordlander projection together with the relations~\eqref{eq-g<-decomposition}, \eqref{eq-gr-decomposition}, \eqref{eq-sigma-decomposition} and~\eqref{eq-sigma-initial} the set of Dyson equations~\eqref{eq-plain-dyson} takes the following form
\begin{widetext}
\begin{subequations}
\begin{align}
\begin{split}
  \frac{\partial}{\partial t} \widetilde{G}_-^<(t,t^\prime) & = - \int_{-\infty}^t\mathrm{d}t_1 \; \widetilde{\sigma}_{\vec{q}}^>(t,t_1) \, b_g(t,t_1) \, \widetilde{G}_-^<(t_1,t^\prime) + \int_{-\infty}^{t^\prime}\mathrm{d}t_1 \; \widetilde{\sigma}_{\vec{k}}^<(t,t_1) \, \widetilde{B}_*^<(t,t_1) \, g_-(t_1,t^\prime) \;,
\end{split} \label{eq-projected-dyson-first} \tag{\theequation a} \\[1ex]
\begin{split}
  \frac{\partial}{\partial t} \widetilde{B}_*^<(t,t^\prime) & = - \int_{-\infty}^t\mathrm{d}t_1 \; \Bigl[ \widetilde{\sigma}_{\vec{k}}^<(t_1,t) \, g_-(t,t_1) \, \widetilde{B}_*^<(t_1,t^\prime) + i \widetilde{\sigma}_{\vec{k}\vec{q}}^>(t,t_1) \, b_g(t,t_1) \, \widetilde{B}_*^<(t_1,t^\prime) \Bigr] \;,
\end{split} \tag{\theequation b} \\[1ex]
\begin{split}
  \frac{\partial}{\partial t} \widetilde{B}_g^<(t,t^\prime) & = \int_{-\infty}^{t^\prime}\mathrm{d}t_1 \; \Bigl[ \widetilde{\sigma}_{\vec{q}}^>(t_1,t) \, \widetilde{G}_-^<(t,t_1) \, b_{g}(t_1,t^\prime) + i \widetilde{\sigma}_{\vec{k}\vec{q}}^>(t_1,t) \, \widetilde{B}_*^<(t,t_1) \, b_g(t_1,t^\prime) \Bigr] \;,
\end{split} \tag{\theequation c} \\[1ex]
\begin{split}
  \Theta(t-t^\prime) \frac{\partial}{\partial t} g_-(t, t^\prime) & = - \int_{t^\prime}^t\mathrm{d}t_1 \; \widetilde{\sigma}_{\vec{q}}^>(t,t_1) \, b_g(t,t_1) \, g_-(t_1,t^\prime) \;,
\end{split} \tag{\theequation d} \\[1ex]
\begin{split}
  \Theta(t-t^\prime) \frac{\partial}{\partial t} b_*(t, t^\prime) & = - \int_{t^\prime}^t\mathrm{d}t_1 \; \Bigl[ \widetilde{\sigma}_{\vec{k}/\vec{q}}^<(t_1,t) \, g_-(t,t_1) \, b_{*/g}(t_1,t^\prime) + i \widetilde{\sigma}_{\vec{k}\vec{q}}^>(t,t_1) \, b_g(t,t_1) \, b_*(t_1,t^\prime) \Bigr] \;,
\end{split} \tag{\theequation e} \\[1ex]
\begin{split}
  \Theta(t-t^\prime) \frac{\partial}{\partial t} b_g(t, t^\prime) & = 0 \;.
\end{split} \label{eq-projected-dyson-last} \tag{\theequation f}
\end{align}\label{eq-projected-dyson}%
\end{subequations}
\end{widetext}

The time evolution of the molecular occupation numbers~$n_-$, $n_*$ and~$n_g$ can be calculated from the following 
relations~\cite{Langreth1991Derivation}
\begin{subequations}
\begin{align}
  \frac{\mathrm{d}n_-(t)}{\mathrm{d}t} & = \frac {\partial \widetilde{G}_-^<(t,t^\prime)}{\partial t} \biggr|_{t=t^\prime} + \frac {\partial \widetilde{G}_-^<(t,t^\prime)}{\partial t^\prime} \biggr|_{t=t^\prime} \;, \\[1ex]
  \frac{\mathrm{d}n_{*}(t)}{\mathrm{d}t} & = \frac {\partial \widetilde{B}_{*}^<(t,t^\prime)}{\partial t} \biggr|_{t=t^\prime} + \frac {\partial \widetilde{B}_{*}^<(t,t^\prime)}{\partial t^\prime} \biggr|_{t=t^\prime} \;, \\[1ex]
  \frac{\mathrm{d}n_{g}(t)}{\mathrm{d}t} & = \frac {\partial \widetilde{B}_{g}^<(t,t^\prime)}{\partial t} \biggr|_{t=t^\prime} + \frac {\partial \widetilde{B}_{g}^<(t,t^\prime)}{\partial t^\prime} \biggr|_{t=t^\prime} \;.
\end{align}\label{eq-occupancies}%
\end{subequations}
Hence, we also need the adjoint Dyson equations of the lesser functions which can be calculated in the same manner. The result reads
\begin{subequations}
\begin{align}
\begin{split}
  \frac{\partial}{\partial t^\prime} \widetilde{G}_-^<(t,t^\prime) & = - \int_{-\infty}^{t^\prime}\mathrm{d}t_1 \; \widetilde{G}_-^<(t,t_1) \, \widetilde{\sigma}_{\vec{q}}^>(t_1,t^\prime) \, b_g(t_1,t^\prime)\\
& + \int_{-\infty}^t\mathrm{d}t_1 \; g_-(t,t_1) \, \widetilde{\sigma}_{\vec{k}}^<(t_1,t^\prime) \, \widetilde{B}_*^<(t_1,t^\prime) \;,
\end{split} \label{eq proj adj dyson G-} \\[1ex]
\begin{split}
  \frac{\partial}{\partial t^\prime} \widetilde{B}_*^<(t,t^\prime) & = - \int_{-\infty}^{t^\prime}\mathrm{d}t_1 \; \widetilde{B}_*^<(t,t_1) \, \widetilde{\sigma}_{\vec{k}}^<(t^\prime,t_1) \, g_-(t_1,t^\prime) \\
  & + - \int_{-\infty}^{t^\prime}\mathrm{d}t_1 \; \widetilde{B}_*^<(t,t_1) \, i \widetilde{\sigma}_{\vec{k}\vec{q}}^>(t_1,t^\prime) \, b_g(t_1,t^\prime) \;,
\end{split} \\[1ex]
\begin{split}
  \frac{\partial}{\partial t^\prime} \widetilde{B}_g^<(t,t^\prime) & = \int_{-\infty}^t\mathrm{d}t_1 \; b_g(t,t_1) \, \widetilde{\sigma}_{\vec{q}}^>(t^\prime, t_1) \, \widetilde{G}_-^<(t_1,t^\prime)\\ 
& + \int_{-\infty}^t\mathrm{d}t_1 b_g(t,t_1) \, i \widetilde{\sigma}_{\vec{k}\vec{q}}^>(t^\prime,t_1) \, \widetilde{B}_*^<(t_1,t^\prime) \;.
\end{split}
\end{align}\label{eq-projected-adjoint-dyson}%
\end{subequations}

Equations~\eqref{eq-projected-dyson} and~\eqref{eq-projected-adjoint-dyson} constitute the final projected set of Dyson equations 
that determines the dynamics of the system within the subspace~$Q=1$. The rate-equation-like structure of these equations is 
already evident. The Dyson equation for the lesser Green function of the negative ion, Eq.~\eqref{eq proj adj dyson G-},
for instance, contains a production term proportional to~$\widetilde{\sigma}_{\vec{k}}$ and~$\widetilde{B}_*^<$ and a loss term 
proportional to~$\widetilde{\sigma}_{\vec{q}}$ and $\widetilde{G}_-^<$. These terms relate to the production and loss of 
negative ions by the RCT electron capture and release reaction, respectively.

Since the self-energies are known in terms of the Green functions, Eqs.~\eqref{eq-projected-dyson} 
and~\eqref{eq-projected-adjoint-dyson} constitute a closed set of equations. A numerical solution along 
the lines pioneered by \citeauthor{Shao1994Manybody}~\cite{Shao1994Manybody} could thus be attempted. The 
rather involved numerics of double-time Green functions is however not required for moderate projectile 
velocities~\cite{Shao1994Manybody}. In that case the semi-classical approximation described in the next 
section can be employed to reduce Eqs.~\eqref{eq-projected-dyson} and~\eqref{eq-projected-adjoint-dyson} 
to a set of rate equations. As far as possible applications of our approach to plasma walls are concerned, 
we have to keep in mind however that plasma walls are usually negatively charged with respect to the bulk 
plasma. Charged projectiles might thus acquire kinetic energies for which the semi-classical approximation 
and with it the rate equations are no longer valid. The metastable molecules however we are concerned with 
in the present work approach the surface with thermal energies making the rate equations an excellent 
approximation to the full two-time equations.

\section{Semi-classical approximation\label{sec semiclassical}}

The strongly oscillating factors of the projected set of Dyson equations are contained in the 
functions~$\widetilde{\sigma}_{\vec{k}}^<$, $\widetilde{\sigma}_{\vec{q}}^>$, and $\widetilde{\sigma}_{\vec{k}\vec{q}}^>$. 
If these functions are strongly peaked along the time diagonal~$t=t^\prime$, we can apply a saddle point approximation 
to the integrals in~\eqref{eq-projected-dyson} and~\eqref{eq-projected-adjoint-dyson}. For instance,
\begin{equation}
\begin{split}
  & \int_{-\infty}^{t^\prime}\mathrm{d}t_1 \; \widetilde{\sigma}_{\vec{k}}^<(t,t_1) \, \widetilde{B}_*^<(t,t_1) \, g_-(t_1,t^\prime) \\
  & \quad \approx \widetilde{B}_*^<(t,t) \, g_-(t,t^\prime) \int_{-\infty}^{t^\prime}\mathrm{d}t_1 \; \widetilde{\sigma}_{\vec{k}}^<(t,t_1) \;.
\end{split}\label{eq semiclassical example}%
\end{equation}\\
The validity of this approximation, which is also known as the semi-classical 
approximation~\cite{Langreth1991Derivation}, will be demonstrated in Sec.~\ref{sec results}. 

Within the saddle-approximation the projected Dyson equations for the lesser Green functions become
\begin{widetext}
\begin{subequations}
\begin{align}
\begin{split}
  \frac{\partial}{\partial t} \widetilde{G}_-^<(t,t^\prime) & \approx - \underbrace{b_g(t,t)}_{1} \widetilde{G}_-^<(t,t^\prime) \int_{-\infty}^t\mathrm{d}t_1 \; \widetilde{\sigma}_{\vec{q}}^>(t,t_1)  + \widetilde{B}_*^<(t,t) \, g_-(t,t^\prime) \int_{-\infty}^{t^\prime}\mathrm{d}t_1 \; \widetilde{\sigma}_{\vec{k}}^<(t,t_1) \;,
\end{split} \\[1ex]
\begin{split}
  \frac{\partial}{\partial t^\prime} \widetilde{G}_-^<(t,t^\prime) & \approx - \widetilde{G}_-^<(t,t^\prime) \underbrace{b_g(t^\prime,t^\prime)}_{1} \int_{-\infty}^{t^\prime}\mathrm{d}t_1 \; \widetilde{\sigma}_{\vec{q}}^>(t_1,t^\prime)  + g_-(t,t^\prime) \, \widetilde{B}_*^<(t^\prime,t^\prime) \int_{-\infty}^t\mathrm{d}t_1 \; \widetilde{\sigma}_{\vec{k}}^<(t_1,t^\prime) \;,
\end{split} \\[1ex]
\begin{split}
  \frac{\partial}{\partial t} \widetilde{B}_*^<(t,t^\prime) & \approx - \underbrace{g_-(t,t)}_{1} \widetilde{B}_*^<(t,t^\prime) \int_{-\infty}^t\mathrm{d}t_1 \; \widetilde{\sigma}_{\vec{k}}^<(t_1,t) - \underbrace{b_g(t,t)}_{1} \widetilde{B}_*^<(t,t^\prime) \int_{-\infty}^t\mathrm{d}t_1 \; i \widetilde{\sigma}_{\vec{k}\vec{q}}^>(t,t_1) \;,
\end{split} \\[1ex]
\begin{split}
  \frac{\partial}{\partial t^\prime} \widetilde{B}_*^<(t,t^\prime) & \approx - \widetilde{B}_*^<(t,t^\prime) \underbrace{g_-(t^\prime,t^\prime)}_{1} \int_{-\infty}^{t^\prime}\mathrm{d}t_1 \; \widetilde{\sigma}_{\vec{k}}^<(t^\prime,t_1) - \widetilde{B}_*^<(t,t^\prime) \underbrace{b_g(t^\prime,t^\prime)}_{1} \int_{-\infty}^{t^\prime}\mathrm{d}t_1 \; i \widetilde{\sigma}_{\vec{k}\vec{q}}^>(t_1,t^\prime) \;,
\end{split} \\[1ex]
\begin{split}
  \frac{\partial}{\partial t} \widetilde{B}_g^<(t,t^\prime) & \approx \widetilde{G}_-^<(t,t) \, b_{g}(t,t^\prime) \int_{-\infty}^{t^\prime}\mathrm{d}t_1 \; \widetilde{\sigma}_{\vec{q}}^>(t_1,t) + \widetilde{B}_*^<(t,t) \, b_{g}(t,t^\prime) \int_{-\infty}^{t^\prime}\mathrm{d}t_1 \; i \widetilde{\sigma}_{\vec{k}\vec{q}}^>(t_1,t) \;,
\end{split} \\[1ex]
\begin{split}
  \frac{\partial}{\partial t^\prime} \widetilde{B}_g^<(t,t^\prime) & \approx b_g(t,t^\prime) \, \widetilde{G}_-^<(t^\prime,t^\prime) \int_{-\infty}^t\mathrm{d}t_1 \; \widetilde{\sigma}_{\vec{q}}^>(t^\prime, t_1) + b_g(t,t^\prime) \, \widetilde{B}_*^<(t^\prime,t^\prime) \int_{-\infty}^t\mathrm{d}t_1 \; i \widetilde{\sigma}_{\vec{k}\vec{q}}^>(t^\prime, t_1) \;.
\end{split}
\end{align}
\end{subequations}
\end{widetext}
Using Eqs.~\eqref{eq-occupancies} we then arrive at a set of rate equations for the occupancies of the molecular 
pseudo-particle states, 
\begin{subequations}
\begin{align}
  \frac{\mathrm{d}n_-(t)}{\mathrm{d}t} & \approx - \Gamma_1(t) \, n_-(t) + \Gamma_0(t) \, n_*(t) \;, \label{eq ode n-} \\
  \frac{\mathrm{d}n_*(t)}{\mathrm{d}t} & \approx - \Gamma_0(t) \, n_*(t) - \Gamma_A(t) \, n_*(t) \;, \label{eq ode n*} \\
  \frac{\mathrm{d}n_g(t)}{\mathrm{d}t} & \approx \Gamma_1(t) \, n_-(t) + \Gamma_A(t) \, n_*(t) \;, \label{eq ode ng}
\end{align}\label{eq-rate-lowest}%
\end{subequations}
where the rates are given by 
\begin{subequations}
\begin{align}
  \Gamma_0(t) & = \int_{-\infty}^{t}\mathrm{d}t_1 \; 2 \Re \bigl\{ \widetilde{\sigma}_{\vec{k}}^<(t,t_1) \bigr\} \;, \label{eq semi rate 0}\\
  \Gamma_1(t) & = \int_{-\infty}^t\mathrm{d}t_1 \; 2 \Re \bigl\{ \widetilde{\sigma}_{\vec{q}}^>(t,t_1) \bigr\} + \Gamma_n \;, \label{eq semi rate 1} \\
  \Gamma_A(t) & = \int_{-\infty}^t\mathrm{d}t_1 \; 2 \Re \bigl\{  i \widetilde{\sigma}_{\vec{k}\vec{q}}^>(t,t_1) \bigr\} \;. \label{eq semi rate auger}
\end{align}\label{eq semi-classical rates}%
\end{subequations}
Note, in Eq.~\eqref{eq semi rate 1} we incorporated the natural decay of the negative ion by adding the natural 
decay rate~$\Gamma_n=1/\tau_n$ on the right-hand side. 

Similar to what Langreth and coworkers did in the context of the neutralization of atomic 
ions~\cite{Langreth1991Derivation,Shao1994Manybody} we have thus reduced a complicated set of Dyson 
equations - Eqs.~\eqref{eq-projected-dyson} and~\eqref{eq-projected-adjoint-dyson} describing the de-excitation 
of a metastable molecule via the simultaneous action of the RCT channel \eqref{eq react rct} and the Auger 
channel \eqref{eq react auger} - to an easy to handle system of rate equations~\eqref{eq-rate-lowest}. The 
reaction rates~\eqref{eq semi-classical rates} entering the rate equations are linked to quantum-kinetic quantities 
and thus related to the semi-empirical model.

The rates~$\Gamma_0$ and~$\Gamma_1$ defined in \eqref{eq semi rate 0} and \eqref{eq semi rate 1} are equal 
to the rates employed in Ref.~\onlinecite{Marbach2012Resonant}. Mathematical expressions for these two
rates - as obtained from the semi-empirical model - can thus be looked up in our previous 
work~\cite{Marbach2012Resonant}. The Auger rate~$\Gamma_A$ introduced in \eqref{eq semi rate auger} however 
has not been calculated before. Within the semi-empirical model it is given by 
\begin{equation}
\begin{split}
  & \Gamma_A(t) = \int_0^\infty \mathrm{d}q_r \, 2 \Re \Biggl\{ \frac{\bigl[ \Gamma_{\vec{k}\vec{q}}(t) \bigr]^*}{\hbar^2 (\Delta q)^3 (\Delta k)^3} \int_{t_0}^t \mathrm{d}t_1 \\
  & \qquad \times \int_0^{\frac{\pi}{2}} \mathrm{d}q_\vartheta \int_0^{2\pi} \mathrm{d}q_\varphi \int_0^{k_F} \mathrm{d}k_r \int_0^\pi \mathrm{d}k_\vartheta \int_0^{2\pi} \mathrm{d}k_\varphi \\
  & \qquad \times q_r^2 \sin(q_\vartheta) \, k_r^2 \sin(k_\vartheta) \Gamma_{\vec{k}\vec{q}}(t_1) \Biggr\} \;
\end{split}\label{eq gamma auger spherical}
\end{equation}
with 
\begin{equation}
  \Gamma_{\vec{k}\vec{q}}(t) = \DPDME(t) \, e^{\frac{i}{\hbar} \int_0^t \mathrm{d}t_1 [ \varepsilon_0(t_1) + \varepsilon_{\vec{q}}(t_1) - \varepsilon_1(t_1) - \varepsilon_{\vec{k}} ]} \;.
\end{equation}

For a given Auger matrix element $\DPDME(t)$ the multi-dimensional integral in~\eqref{eq gamma auger spherical} 
can be calculated efficiently using the techniques and approximations outlined in Ref.~\onlinecite{Marbach2011Auger}. 
The Auger matrix element, originating from the Coulomb interaction between the (active) projectile electron and 
an electron from the solid, is in general subject to the dynamical response of the target electrons. For metallic
surfaces this is an important issue, as discussed for instance by \citeauthor{Alvarez1998Auger}~\cite{Alvarez1998Auger}. 
It leads to the screening of the Coulomb interaction and should be at least accounted for by a statically 
screened Coulomb potential. For the dielectric surfaces we are primarily interest in, however, screening is suppressed 
by the energy gap. We calculate therefore $\DPDME(t)$ from the bare Coulomb interaction. Thereby we overestimate 
somewhat the strength of the Auger matrix element. 

We will now seek an analytic solution to the coupled rate equations~\eqref{eq-rate-lowest}. As a starting point we first 
take a step back and consider the isolated decay channels of resonant electron capture, resonant electron emission, and Auger 
de-excitation. Singling out the individual reactions in~\eqref{eq-rate-lowest} we obtain
\begin{subequations}
\begin{align}
  \frac{\mathrm{d}n_*^{(0)}(t)}{\mathrm{d}t} & = -\Gamma_0(t) \, n_*^{(0)}(t) \;, \label{eq rate * isolated}\\
  \frac{\mathrm{d}n_-^{(1)}(t)}{\mathrm{d}t} & = -\Gamma_1(t) \, n_-^{(1)}(t) \;, \\
  \frac{\mathrm{d}n_*^{(A)}(t)}{\mathrm{d}t} & = -\Gamma_A(t) \, n_*^{(A)}(t) \;.
\end{align}\label{eq isolated system}%
\end{subequations}

The superscripts~${(0)}$, ${(1)}$, and~${(A)}$ identify the isolated resonant electron capture, resonant electron emission,
and Auger de-excitation, respectively. Since the channels are isolated, each of the decay equations 
\eqref{eq isolated system} comes with an analogous 
equation for the species that is produced. For instance, accompanying~\eqref{eq rate * isolated} is the equation
\begin{equation}
  \frac{\mathrm{d}n_-^{(0)}(t)}{\mathrm{d}t} = \Gamma_0(t) \, n_*^{(0)}(t) \;.
  \label{eq rate - isolated}
\end{equation}
The time derivatives of~$n_*^{(0)}$ and~$n_-^{(0)}$ differ however only in sign. Hence,~$n_-^{(0)}$ is given through the 
conservation of particles as~${n_-^{(0)} = 1 - n_*^{(0)}}$ (valid when the channels are isolated). Consequently, the 
additional equations of type~\eqref{eq rate - isolated} do not contain any additional information and can be omitted. Using 
the initial condition~${n_*^{(0)}(t_0)=n_*^{(A)}(t_0)=1}$ the system~\eqref{eq isolated system} can be solved 
straightforwardly. The result is
\begin{subequations}
\begin{align}
  n_*^{(0)}(t) & = e^{-\int_{t_0}^t \mathrm{d}t_1 \, \Gamma_0(t)} \;, \\
  n_-^{(1)}(t) & = n_-^{(1)}(t^\prime) \, e^{-\int_{t^\prime}^t \mathrm{d}t_1 \, \Gamma_1(t)} \;, \\
  n_*^{(A)}(t) & = e^{-\int_{t_0}^t \mathrm{d}t_1 \, \Gamma_A(t)} \;.
\end{align}\label{eq isolated solutions}%
\end{subequations}

Now we are in position to use these occupancies to calculate the solution of the full, coupled system 
of rate equations~\eqref{eq-rate-lowest}. First we consider the equation for $n_*$. Using the initial 
condition ${n_*(t_0)=1}$, Eq.~\eqref{eq ode n*} can be solved by separation of variables and yields
\begin{equation}
  n_*(t) = e^{-\int_{t_0}^t \mathrm{d}t_1 \bigl( \Gamma_0(t_1) 
  + \Gamma_1(t_1) \bigr)} = n_*^{(0)}(t) \, n_*^{(A)}(t) \;. \label{eq n* solution}
\end{equation}

To solve~\eqref{eq ode n-} for the occupancy of the negative ion state we first multiply this equation by a 
factor ${\exp\bigl(\int_{t_0}^t \mathrm{d}t_2 \, \Gamma_1(t_2)\bigr)}$ and rearrange the terms to obtain
\begin{equation}
  \frac{\mathrm{d}}{\mathrm{d}t} \biggl( n_-(t) \, e^{\int_{t_0}^t \mathrm{d}t_2 \, \Gamma_1(t_2)} \biggr) = \Gamma_0(t) \, n_*(t) \, e^{\int_{t_0}^t \mathrm{d}t_2 \, \Gamma_1(t_2)} \;.
\end{equation}
Relabeling then $t$ as $t_1$ and integrating the equation from ${t_1=t_0}$ to ${t_1=t}$ while minding the initial 
condition ${n_-(t_0)=0}$ yields after a further rearrangement
\begin{equation}
\begin{split}
  n_-(t) & = \int_{t_0}^t \mathrm{d}t_1 \, \Gamma_0(t_1) \, n_*(t_1) \, e^{-\int_{t_1}^t \mathrm{d}t_2 \, \Gamma_1(t_2)} \\
  & = \int_{t_0}^t \mathrm{d}t_1 \biggl[ - \frac{\mathrm{d}n_*^{(0)}(t_1)}{\mathrm{d}t_1} \biggr] n_*^{(A)}(t_1) \, \frac{n_-^{(1)}(t)}{n_-^{(1)}(t_1)} \;.
\end{split}\label{eq n- solution}%
\end{equation}

Finally, the occupancy of the molecular ground state $n_g$, that is, the solution of Eq.~\eqref{eq ode ng}, is 
given through the particle conservation property of the full system~\eqref{eq-rate-lowest}, 
\begin{equation}
  n_g(t) = 1- n_*(t) - n_-(t) \;. \label{eq ng solution}
\end{equation}

Note, the molecular occupancies satisfying the combined rate equation scheme Eqs.~\eqref{eq n* solution}, 
\eqref{eq n- solution}, and \eqref{eq ng solution} are completely determined by the occupancies 
$n_*^{(0)}$, $n_-^{(1)}$ and $n_*^{(A)}$. Moreover, when the Auger channel is disabled by 
setting~${\Gamma_A(t) \equiv 0}$, Eqs.~\eqref{eq n* solution}, \eqref{eq n- solution} and \eqref{eq ng solution} 
reduce to the rate equations derived by intuitive means for the isolated RCT channel~\cite{Marbach2012Resonant}. 
Hence, the quantum-kinetic treatment justifies a posteriori the intuitive approach taken 
by us in Ref.~\onlinecite{Marbach2012Resonant}.

We now turn to the spectrum of the emitted electron. While the evolution of the~$\vec{q}$ states has not been 
considered explicitly in our quantum-kinetic calculation, the occupancy of these states can nevertheless be 
extracted from the solution of~\eqref{eq-rate-lowest}. 

From the reactions~\eqref{eq react rct} and~\eqref{eq react auger} it is obvious that the probability for 
emitting an electron $n_{e}(t)$ is equal to the occupancy of the ground state $n_g(t)$ as every ground state 
molecule must have resulted from the reaction chain and, hence, must be accompanied by an emitted electron. 
Consequently, the evolution of~$n_{e}(t)$ is governed by Eq.~\eqref{eq ode ng}. Due to the image potential, however, 
not every emitted electron can escape the surface. In particular, the escape is only possible when the emitted 
electron's perpendicular kinetic energy~${\varepsilon_{q_z}^\infty=\varepsilon_{\vec{q}}^\infty \cos^2(q_\vartheta)}$ 
is higher than the absolute value of the image potential~$V_i$ at the position of emission. The latter can be 
approximated by the position of the molecule's center of mass at the time of emission. 

To incorporate the image potential effect, we adopt a two step strategy. As a start we introduce the spectral 
rates $\varrho_1(\varepsilon_{\vec{q}}^\infty,t)$ and $\varrho_A(\varepsilon_{\vec{q}}^\infty,t)$ which are not 
restricted by the image potential effect by writing
\begin{equation}
  \Gamma_{1/A}(t) = \int_0^\infty \mathrm{d}\varepsilon_{\vec{q}}^\infty \, \varrho_{1/A}(\varepsilon_{\vec{q}}^\infty,t) \;, \label{eq spectral rate defs}
\end{equation}
and afterwards we let~$\varrho_{1/A}\rightarrow\bar{\varrho}_{1/A}$ with
\begin{equation}
\begin{split}
  & \bar{\varrho}_{1/A}(\varepsilon_{\vec{q}}^\infty,t) = \int_0^{\frac{\pi}{2}} \mathrm{d}q_\vartheta \int_0^{2\pi} \mathrm{d}q_\varphi \\\
  & \qquad \times \Theta\bigl( V_i(z_R(t)) + \varepsilon_{q_z}^\infty \bigr) \, \frac{\mathrm{d}^2 \varrho_{1/A}(\varepsilon_{\vec{q}}^\infty,t)}{\mathrm{d}q_\vartheta \mathrm{d}q_\varphi} \;.
\end{split}\label{eq spectral cut-off}%
\end{equation}

An explicit expression for the spectral RCT emission rate~$\varrho_1$ has been given in Ref.~\onlinecite{Marbach2012Resonant}. 
The spectral Auger rate~$\varrho_A$ may be calculated from Eq.~\eqref{eq gamma auger spherical} by stripping out the 
$q_r$~integral and multiplying the result by~$m_e/\hbar^2 q_r$. Introducing the spectral decomposition of the rates 
in Eq.~\eqref{eq ode ng} and identifying $n_g$ with $\bar{n}_{e}$ we obtain
\begin{equation}
\begin{split}
  \frac{\mathrm{d}\bar{n}_{}(t)}{\mathrm{d}t} & = \int_0^\infty \mathrm{d}\varepsilon_{\vec{q}}^\infty \, \bar{\varrho}_1(\varepsilon_{\vec{q}}^\infty,t) \, n_-(t) \\
  & \phantom{=} + \int_0^\infty \mathrm{d}\varepsilon_{\vec{q}}^\infty \, \bar{\varrho}_A(\varepsilon_{\vec{q}}^\infty,t) \, n_*(t) \;,
\end{split}
\end{equation}
where $\bar{n}_{e}$ denotes the probability for emitting an electron that can escape from the surface. Integrating over 
the time argument with the initial condition ${\bar{n}_{e}(t_0)=0}$ and taking the derivative with respect 
to~$\varepsilon_{\vec{q}}^\infty$ we find for the spectrum of the emitted electron at time~$t$
\begin{equation}
\begin{split}
  \frac{\mathrm{d}\bar{n}_{e}}{\mathrm{d}\varepsilon_{\vec{q}}^\infty} \biggr|_t & = \int_{t_0}^t \mathrm{d}t_1 \, \bar{\varrho}_1(\varepsilon_{\vec{q}}^\infty,t_1) \, n_-(t_1) \\
  & \phantom{=} + \int_{t_0}^t \mathrm{d}t_1 \, \bar{\varrho}_A(\varepsilon_{\vec{q}}^\infty,t_1) \, n_*(t_1) \;.
\end{split}\label{eq spectrum}%
\end{equation}
The secondary electron emission coefficient~$\gamma_e$, that is, the probability for having emitted an electron 
after the collision is completed, is given by 
\begin{equation}
  \gamma_e = \bar{n}_{e}(\infty) \;. \label{eq gamma}
\end{equation}

The occupancies of the molecular pseudo-particle states \eqref{eq n* solution}, \eqref{eq n- solution}, 
\eqref{eq ng solution} and the spectrum of the emitted electron \eqref{eq spectrum} are the main result 
of this work. The occupancies fully characterize the temporal evolution of the de-excitation of a 
metastable \NitrogenDominantMetastableState\ molecule at a surface when both the RCT and the 
Auger channel are open. The ingredients required as an input, the occupancies arising from the isolated 
processes~$n_*^{(0)}$, $n_-^{(1)}$, $n_*^{(A)}$ and the image potential adjusted spectral 
rates~$\bar{\varrho}_1$, $\bar{\varrho}_A$, can be obtained from the quantum-kinetic calculation 
and thus from the semi-empirical model. 

Assuming the parameters of the model Hamiltonian to be a priori fixed, either by 
experiment or by quantum-chemical calculations, there is no free parameter in the kinetic equations
which can be a posteriori adjusted to experimental data concerning the surface collision itself. This is 
in contrast to Hagstrum's theory of secondary electron emission~\cite{Hagstrum1954Auger,Hagstrum1961Theory} 
where the matrix elements (more precisely the combined density of states) are directly fitted to the outcome 
of the surface scattering experiment.

\section{Numerical results\label{sec results}}

\begin{figure}
  \centerline{\includegraphics[scale=1]{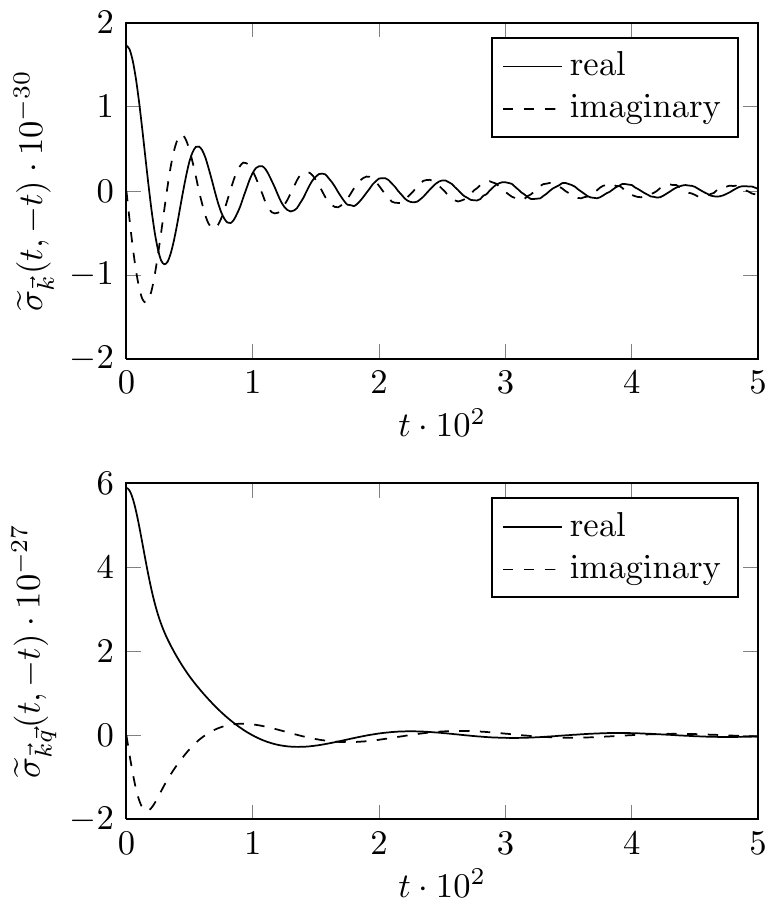}}
  \caption{Variation of the real (solid line) and imaginary (dashed line) part of~${\widetilde{\sigma}_{\vec{k}}(t_1,t_2)}$ 
(upper panel) and~${\widetilde{\sigma}_{\vec{k}\vec{q}}(t_1,t_2)}$ (lower panel) as calculated from~\eqref{eq sigma definitions} 
along the anti-diagonal~${t_1=-t_2=t}$. The molecule's axis was aligned perpendicular to the surface and the molecule's kinetic 
energy was fixed to~${50\,meV}$. The behavior for negative time arguments~$t$ is omitted since the real (imaginary) part of both 
functions is symmetric (anti-symmetric) with respect to the time diagonal. Note that the time~$t$ is dimensionless. It relates 
to the physical time~$t_{\rm phys}$ via~$t_{\rm phys}=a_B \Delta t /2 v_0$.\label{fig sigma k kq}}
\end{figure}

In this section we present numerical results based on the semi-classical equations of Sec.~\ref{sec semiclassical}. 
We consider the particular case of a diamond surface and restrict our investigations to normal incident with a 
molecular kinetic energy of~${50\,meV}$. The turning point of the molecule's trajectory is then $4.4$ Bohr 
radii~\cite{Katz1995Temperature}. As in our previous work~\cite{Marbach2011Auger,Marbach2012Resonant} we treat 
only the two principal orientations of the metastable \NitrogenDominantMetastableState\ molecule: molecular 
axis perpendicular to the surface and molecule axis parallel to the surface. Furthermore, we omit the surface 
induced decay channel by setting~${\Gamma_1(t)=\Gamma_n}$. As our previous investigations showed, this is an 
excellent approximation~\cite{Marbach2012Resonant}.
\begin{figure}
  \centerline{\includegraphics[scale=1]{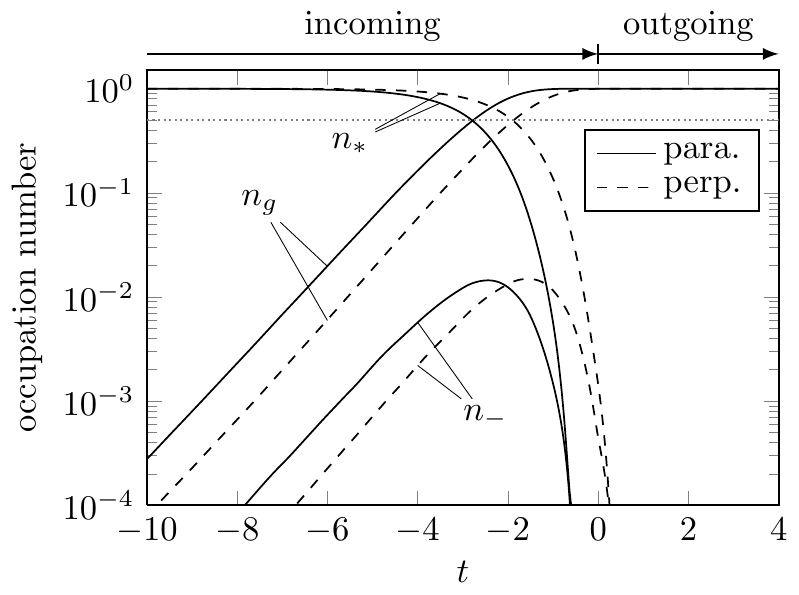}}
  \caption{Time dependence of the occupancy of the ground state molecule~$n_g$, the metastable molecule~$n_*$, and 
  the negative ion~$n_-$ in the parallel (solid line) and perpendicular (dashed line) orientation. The molecule's 
  kinetic energy was fixed to~${50\,meV}$. The dotted line represents half filling of the respective level. 
  The $y$-axis is logarithmic and the time is denoted in the dimensionless units of Fig.~\ref{fig sigma k kq}. 
  The curves were calculated from Eqs.~\eqref{eq n* solution}, \eqref{eq n- solution} and~\eqref{eq ng solution},
  respectively. The incoming and outgoing branches of the trajectory are indicated at the top of the
diagram.\label{fig n2}}
\end{figure}

The numerics necessary to calculate within the semi-empirical model the Auger matrix element $\DPDME$ has been 
described in Ref.~\onlinecite{Marbach2011Auger}. Utilizing the fact that for the low collision energies we 
are interested in the turning point of the molecule is far outside the solid as well as the fact that the LCAO 
wave functions are strongly localized on the molecule whereas the wave functions of the solid and free electron 
are bounded in the mathematical sense we split-off the time-dependence of the matrix element and integrate the
rest by an interpolative grid-based Monte Carlo scheme. The tunneling matrix elements $\RCTCaptureME$ and 
$\RCTSIReleaseME$ can be calculated within the semi-empirical model partly analytically and partly numerically. 
The required tools are given in Ref.~\onlinecite{Marbach2012Resonant}. 

First, we investigate the validity of the semi-classical approximation. As outlined above, approximations of the 
form~\eqref{eq semiclassical example} are only acceptable if the functions~${\widetilde{\sigma}_{\vec{k}}(t_1,t_2)}$ 
and~${\widetilde{\sigma}_{\vec{k}\vec{q}}(t_1,t_2)}$ are sufficiently peaked on the time diagonal~${t_1=t_2}$. In 
order to validate this assumption for our case, we plot in Fig.~\ref{fig sigma k kq} for the perpendicular 
orientation the variation of these two functions along the anti-diagonal~${t_1=-t_2}$. The plots can be generated 
directly from the definitions~\eqref{eq sigma definitions} and represent profiles with respect to the time diagonal.

For both functions ${\widetilde{\sigma}_{\vec{k}}(t_1,t_2)}$ and ${\widetilde{\sigma}_{\vec{k}\vec{q}}(t_1,t_2)}$ 
the real part has its maximum on the time diagonal whereas the imaginary part vanishes on the diagonal 
itself but exhibits the highest value very close to it. When the time separation from the diagonal is enlarged, both 
functions decrease in an oscillating way. For~${\widetilde{\sigma}_{\vec{k}}}$ the amplitude decreases to about~$10\%$ 
over a dimensionless time interval of~${\Delta t \approx 0.05}$. This relates to a physical time 
span of~$\Delta t_{\rm phys} = a_B \Delta t / 2 v_0 \approx 2.25\,fs$ and to the motion of the molecule along 
a distance of~${0.025\,a_B}$. For ${\widetilde{\sigma}_{\vec{k}\vec{q}}}$ the fall-off is even more drastic. The behavior 
for shifted anti-diagonals as well as for parallel orientation is very similar and hence not shown here. Altogether, 
we can conclude that with respect to the macroscopic motion of the molecule the functions~${\widetilde{\sigma}_{\vec{k}}}
$ and~${\widetilde{\sigma}_{\vec{k}\vec{q}}}$ are indeed sufficiently peaked on the time diagonal. Thus, the 
semi-classical approximation is valid in our case.

We now turn to the occupancies of the molecular pseudo-particle states, that is, the probabilities with which the 
molecular configurations involved in the de-excitation process appear in the course of the scattering event. The 
time dependent occupancy of the ground state~$n_g$, the metastable state~$n_*$, and negative ion state~$n_-$ 
can be calculated from Eqs.~\eqref{eq n* solution}, \eqref{eq n- solution} and~\eqref{eq ng solution}, 
respectively. The results are depicted in a semi-logarithmic plot in Fig.~\ref{fig n2}.

Inspection of the curves in Fig.~\ref{fig n2} reveals that even close to the surface the occupancy of the negative 
ion state is rather low. Hence, the metastable projectile is almost immediately converted into a ground state 
molecule and thus stays mostly neutral during the whole collision. 
In Fig.~\ref{fig n2} this fact is recognizable at the crossing point of the~$n_*$ and~$n_g$ curves which occurs 
at approximately half filling of both levels. The low occupancy of the negative ion state is caused by the 
high efficiency of the natural decay channel~\cite{Marbach2012Resonant} and not by the Auger channel destroying 
the metastable molecule, which is the generating species of the negative ion. In fact, it is the other way 
around and in order to substantiate this claim, we investigate below the relative efficiency of the RCT and 
Auger channel by considering the respective reaction rates. Before we do that let us note however that due
the neutrality of the projectile along most of its path it would not gain much kinetic energy in front of a 
charged surface. We expect therefore the semi-classical approximation and hence the rate equations to be 
also valid in case the de-excitation occurred in front of a negatively charged plasma wall.

\begin{figure}
  \centerline{\includegraphics[scale=1]{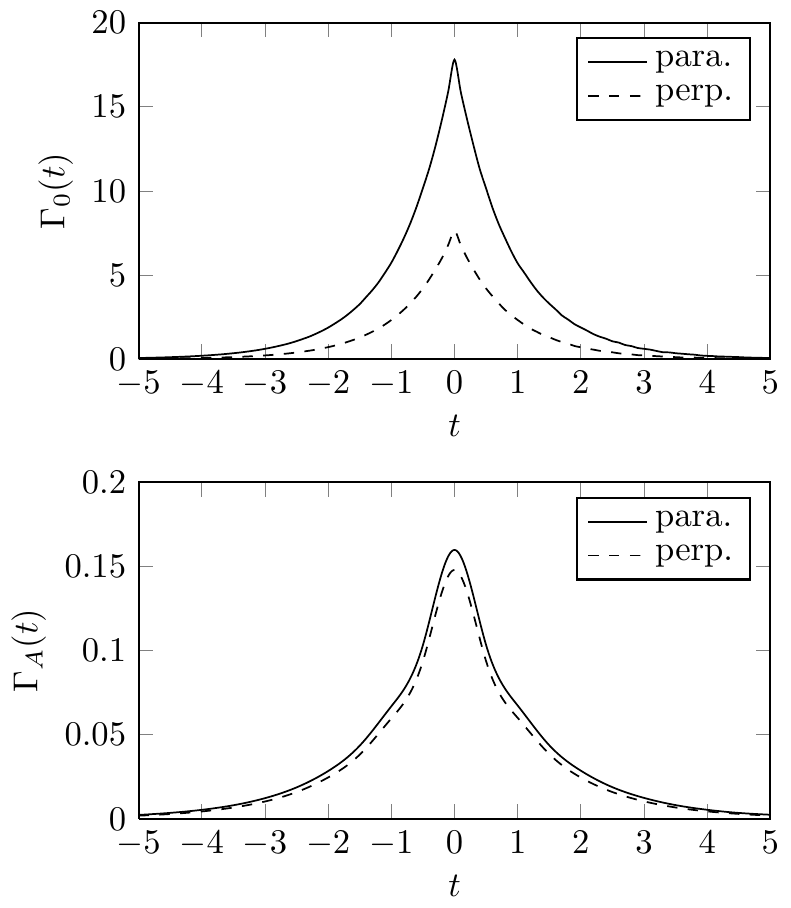}}
  \caption{Variation of the rates of resonant electron capture~$\Gamma_0$ (upper panel) and Auger de-excitation~$\Gamma_A$ 
(lower panel) for the parallel (solid line) and perpendicular (dashed line) orientation at a kinetic energy of~${50\,meV}$. 
The time is denoted in the dimensionless units of Fig.~\ref{fig sigma k kq}.\label{fig gamma 0 a}}
\end{figure}
Figure~\ref{fig gamma 0 a} shows the rates of resonant electron capture~$\Gamma_0$ and Auger de-excitation~$\Gamma_A$.
For both channels the rates are highest at the molecule's turning point (approximately $4.4~a_B$) which is the point 
of smallest molecule-surface separation and strongest molecule-surface interaction. 
When the molecule-surface distance is increased, the rates decrease 
exponentially. The RCT channel's rate is about two orders of magnitude higher than the Auger channel's rate. Consequently, 
the RCT channel captures surface electrons much more efficiently than the Auger channel. In fact, the RCT channel is so 
effective in capturing electrons that it under-runs the Auger channel by destroying its starting basis, the metastable 
state. As a result, in the combined two-channel system the Auger channel's performance is significantly diminished as 
compared to the isolated Auger reaction. 

This conclusion may be verified by considering the term in the rate equations which is responsible for the production 
of the ground state molecule by an Auger de-excitation. It is given by 
(see Eqs.~\eqref{eq ode ng} and~\eqref{eq n* solution})
\begin{equation}
  \Gamma_A(t) \, n_*(t) = \Gamma_A(t) \, n_*^{(A)}(t) \, n_*^{(0)}(t) \;. \label{eq ng auger term}
\end{equation}
Here the factor~${n_*^{(0)}(t)}$ is only present in the combined two-channel system but not in the isolated Auger 
system. Without explicit proof but based on numerical observations we note that the term~${n_*^{(0)}(t)}$ is almost 
identical to the combined 
occupation~${n_*(t)}$ depicted in Fig.~\ref{fig n2}. Hence, in the combined system the Auger channel's ground state 
production term~\eqref{eq ng auger term} is strongly suppressed already in the incoming branch of the trajectory.
\begin{figure}
  \centerline{\includegraphics[scale=1]{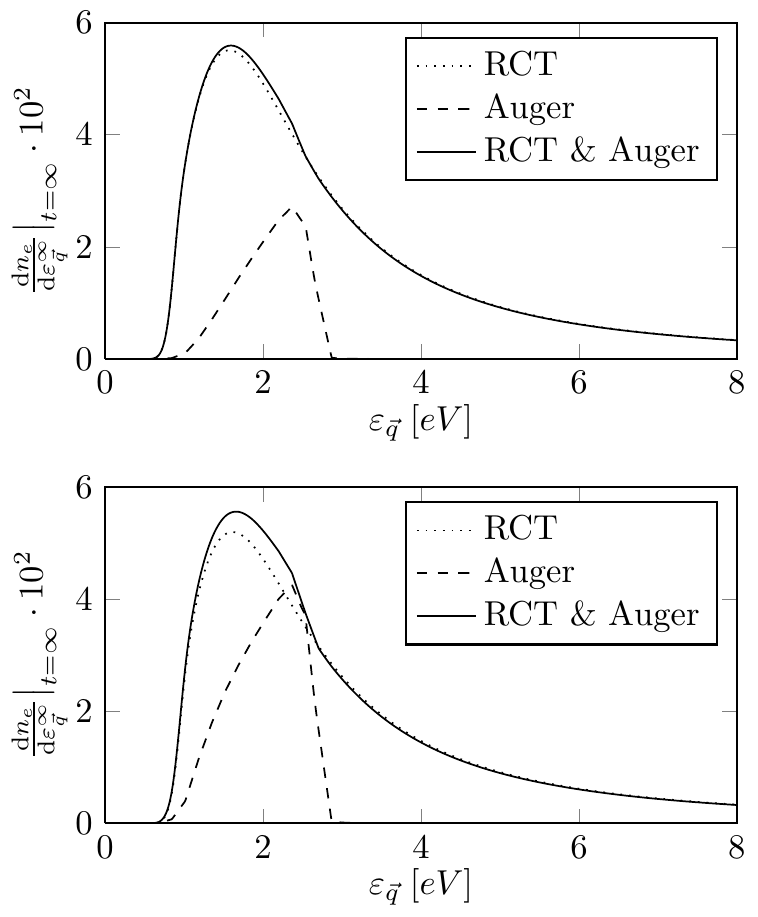}}
  \caption{Energy spectrum of the emitted electron in parallel (upper panel) and perpendicular (lower panel) orientation 
calculated from~\eqref{eq spectrum} at~${t=\infty}$ for a kinetic energy of~$50\,meV$. In both panels the dotted line 
specifies the isolated RCT spectrum (obtained by setting~${\bar{\varrho}_A \equiv 0}$), the dashed line denotes the isolated 
Auger spectrum (obtained by setting~${\bar{\varrho}_1 \equiv 0}$), and the solid line represents the combined 
two-channel spectrum.\label{fig spectrum}}
\end{figure}

Finally, we turn to the energy spectrum of the emitted electron. Figure~\ref{fig spectrum} depicts the 
emission spectrum at $t=\infty$ for the combined two-channel reaction as well as for the isolated reaction 
channels. The latter can be obtained by setting in~\eqref{eq spectrum} ${\bar{\varrho}_A \equiv 0}$ or 
${\bar{\varrho}_1 \equiv 0}$. As can be seen, the isolated RCT spectra exhibit a strong peak at about~$1.5\,eV$ and 
slowly drop off for higher energies. The isolated Auger spectra, on the other hand, monotonously increase until 
approximately~$2.8\,eV$ and then immediately fall off. The low energy cut-off of all curves is due to the 
trapping of the emitted electron in the image potential close to the surface when its perpendicular energy
is too low. The combined spectra almost equal the respective isolated RCT spectra. Only in the range from~${1.5\,eV}$ 
to~${2.5\,eV}$ are the combined spectra slightly increased with respect to the RCT curves. This minor enlargement 
is due to the Auger channel and supports our previous finding that the RCT channel dominates the Auger channel. 

The combined spectra in Fig.~\ref{fig spectrum} are different from the simple addition of the isolated spectra. 
This behavior is caused by the unified treatment of the RCT and Auger reaction channels. The effect would be 
even more pronounced for molecular species forming stable negative ions. Here the resonant electron emission 
would be almost completely blocked as the surface induced decay is always very weak. The resonant electron 
capture, however, would be still very efficient in destroying the initial species. Consequently, the spectrum 
of the emitted electron would resemble the Auger spectrum in shape but would be strongly decreased in magnitude. 

The secondary electron emission coefficients are given by the area beneath the curves in 
Fig.~\ref{fig spectrum} and are summarized in Table~\ref{tab gamma}.
In accordance with our previous observations the emission coefficients are not changed significantly by the inclusion 
of the Auger channel. A similar result was found by \citeauthor{Stracke1998Formation}~\cite{Stracke1998Formation} for 
\NitrogenDominantMetastableState\ de-exciting at a tungsten surface. Their experimental measurements imply that only 
about 10\% of the secondary electron emission coefficient is made up by the Auger channel.

\section{Conclusions\label{sec conclusion}}

We constructed in this work a semi-empirical generalized Anderson-Newns model for secondary electron emission 
due to de-excitation of metastable~\NitrogenDominantMetastableState\ molecules at dielectric surfaces. The 
model treats Auger de-excitation and the two-step resonant charge transfer 
process, where the \NitrogenNegativeIonResonance\ ion acts as a relay state, on an equal footing. 
It reduces the molecular projectile to a two-level system representing the molecular 
orbitals which change their occupancies during the reaction and treats the surface as a simple step 
potential confining the electrons of the solid. 
By construction, the semi-empirical model is not restricted to a particular projectile-target combination. 
Having applications of the model to charge-transferring processes at plasma walls in mind, where 
a great variety of different projectile-target combinations occurs, we consider this as a real advantage. 
Another advantage is that the semi-empirical model separates the  many-body theoretical description of the 
non-interacting projectile and target from the quantum-kinetic treatment of the scattering process. The 
former is simply encapsulated in the parameters of the model Hamiltonian and the latter is performed by 
Green functions. This is particularly advantageous in cases where the surface scattering event is studied
primarily because of its connection to the physics of quantum-impurities.
\begin{table}
  \begin{ruledtabular}
    \begin{tabular}{c|d|d}
      & \text{$\gamma_e^\parallel$} & \text{$\gamma_e^\perp$} \\\hline
      RCT & 0.16685 & 0.15873 \\
      Auger & 0.02760 & 0.04921 \\
      RCT \& Auger & 0.16754 & 0.16335
    \end{tabular}
  \end{ruledtabular}
  \caption{Secondary electron emission coefficients in parallel ($\gamma_e^\parallel$) and perpendicular ($\gamma_e^\perp$)
  orientation at a kinetic energy of~${50\,meV}$.\label{tab gamma}}
\end{table}

For the semi-empirical model to work a method was required to assign and control the energies of the 
two-level system in accordance to the reaction channels, that is, to have the two-level system describing  
all three molecular configurations involved in the de-excitation process: the metastable 
\NitrogenDominantMetastableState\ molecule, the negative ion \NitrogenNegativeIonResonance, and the 
molecular ground state \NitrogenGroundState. We showed how this can be done with projection 
operators and auxiliary bosons. As a result, both the resonant tunneling and the Auger channel could 
be cast into a single model Hamiltonian which, with the help of pseudo-particle operators, could then 
be made amenable to a diagrammatic quantum-kinetic calculation. Using the self-consistent 
non-crossing approximation for the self-energies and a saddle-point approximation for the time integrals 
in the self-energies we finally derived from the Dyson equations for the propagators of the molecular 
pseudo-particles a set of rate equations for the probabilities with which the molecular configurations 
contributing to the de-excitation process can be found in the course of the scattering event. Without 
the Auger channel, the system of rate equations reduces to the one postulated by us before on intuitive 
grounds for the RCT channel alone~\cite{Marbach2012Resonant}. The present work justifies therefore this
reasoning a posteriori. 
%Since the rates are obtained from quantum-kinetic expressions
%they are directly linked to the underlying semi-empirical model. 

For the particular case of a diamond surface we verified the validity of the semi-classical approximation and 
investigated for a collision energy of $50\, meV$ the interplay of the resonant tunneling and the Auger channel. 
In particular, we analyzed the temporal evolution of the probabilities with which the projectile is to be 
found in the \NitrogenDominantMetastableState, the \NitrogenNegativeIonResonance, or the \NitrogenGroundState\
state and explicitly calculated the rates for electron capture due respectively to tunneling and Auger 
de-excitation. 
We also obtained the spectrum of the emitted electron and the secondary electron emission coefficient $\gamma$ 
which are the two quantities of main importance for the modeling of gas discharges. Our results indicate 
for a diamond surface and a kinetic energy of $50\,meV$ the resonant tunneling channel clearly dominating the
Auger channel. The contribution of the Auger channel to the secondary electron emission coefficient lies only in 
the range of a few percent. The overall $\gamma$ coefficient is on the order of $10^{-1}$ in agreement with
what has to be typically assumed to make kinetic simulations of dielectric barrier discharges reproduce the 
properties of the discharge. 

With minor modifications the semi-empirical model and its quantum-kinetic handling leading to the easy
to use set of rate equations can be adopted to other plasma-relevant charge-transferring surface collisions 
as well. At least for low-energy collisions, where the projectile velocities are low enough to allow for a 
reduction of the full double-time kinetic equations to a set of simple rate equations, we can thus hope 
to replace the rules of the thumb which are often needed to characterize secondary electron emission 
due to neutral and charged heavy plasma species hitting the plasma wall by plausible quantitative estimates.
% In all other cases the full double-time 
% kinetic equations have to be solved numerically.}

\begin{acknowledgments}
Johannes Marbach was funded by the federal state of Mecklenburg-Western Pomerania through a postgraduate scholarship. 
In addition this work was supported by the Deutsche Forschungsgemeinschaft through the Transregional Collaborative 
Research Center SFB/TRR24.
\end{acknowledgments}

\appendix

\section{Langreth-Wilkins rules\label{app langreth}}

The Langreth-Wilkins rules~\cite{Langreth1972Theory} are a powerful tool for the analytic continuation of 
propagators defined on a complex time contour onto the real time axis. Their explicit form depends on the 
initial definition of the Green functions. Unfortunately, there is no common agreement about the usage of 
$i$ factors. Moreover, rules published in the past sometimes contained typographic errors~\cite{Langreth1991Derivation}. 
Due to these reasons we list the explicit form of the Langreth-Wilkins rules used in this work.
The rules can be derived in the standard way~\cite{Langreth1972Theory,HaugJauho1996} using however 
the definitions \eqref{contourGF}--\eqref{eq-retarded-definition} for the Green functions. In the following 
$F$ and $B$ denote fermion and boson propagators, respectively. 

To analytically continue the boson-like fermion-antifermion pair
\begin{equation}
  B(t,t^\prime) = F_1(t,t^\prime) \, F_2(t^\prime,t) \;,
\end{equation}
we utilize
\begin{subequations}
\begin{align}
  B^>(t,t^\prime) & = i \, F_1^>(t,t^\prime) \, F_2^<(t^\prime,t) \;, \\
  B^<(t,t^\prime) & = i \, F_1^<(t,t^\prime) \, F_2^>(t^\prime,t) \;, \\
%\begin{split}
%  B^R(t,t^\prime) & = -i \Bigl[ F_1^>(t,t^\prime) \, F_2^A(t^\prime,t) \\
%& \phantom{= -i \Bigl[} + F_1^R(t,t^\prime) \, F_2^>(t^\prime,t) \Bigr]
%\end{split} \\
\begin{split}
  B^R(t,t^\prime) & = i \Bigl[ F_1^<(t,t^\prime) \, F_2^A(t^\prime,t) \\
  & \phantom{= i \Bigl[} + F_1^R(t,t^\prime) \, F_2^<(t^\prime,t) \Bigr] \;,
\end{split} \\
%\begin{split}
%  B^A(t,t^\prime) & = -i \Bigl[ F_1^>(t,t^\prime) \, F_2^R(t^\prime,t) \\
%  & \phantom{= -i \Bigl[} + F_1^A(t,t^\prime) \, F_2^>(t^\prime,t) \Bigr]
%\end{split} \\
\begin{split}
  B^A(t,t^\prime) & = i \Bigl[ F_1^<(t,t^\prime) \, F_2^R(t^\prime,t) \\
  & \phantom{= i \Bigl[} + F_1^A(t,t^\prime) \, F_2^<(t^\prime,t) \Bigr] \;.
\end{split}
\end{align}
\end{subequations}

For the fermion-like fermion-boson pair
\begin{equation}
  F(t,t^\prime) = F_1(t,t^\prime) \, B_1(t,t^\prime) \;,
\end{equation}
the following rules hold
\begin{subequations}
\begin{align}
  F^>(t,t^\prime) & = -i \, F_1^>(t,t^\prime) \, B_1^>(t,t^\prime) \;, \\
  F^<(t,t^\prime) & = -i \, F_1^<(t,t^\prime) \, B_1^<(t,t^\prime) \;, \\
%\begin{split}
%  F^R(t,t^\prime) & = -i \Bigl[ F_1^R(t,t^\prime) \, B_1^>(t,t^\prime) \\
%& \phantom{= -i \Bigl[} - F_1^<(t,t^\prime) \, B_1^R(t,t^\prime) \Bigr]
%\end{split} \\
\begin{split}
  F^R(t,t^\prime) & = -i \Bigl[ F_1^R(t,t^\prime) \, B_1^<(t,t^\prime) \\
  & \phantom{= -i \Bigl[} + F_1^>(t,t^\prime) \, B_1^R(t,t^\prime) \Bigr] \;,
\end{split} \\
%\begin{split}
%  F^A(t,t^\prime) & = -i \Bigl[ F_1^A(t,t^\prime) \, B_1^>(t,t^\prime) \\
%  & \phantom{= -i \Bigl[} - F_1^<(t,t^\prime) \, B_1^A(t,t^\prime) \Bigr]
%\end{split} \\
\begin{split}
  F^A(t,t^\prime) & = -i \Bigl[ F_1^A(t,t^\prime) \, B_1^<(t,t^\prime) \\
  & \phantom{= -i \Bigl[} + F_1^>(t,t^\prime) \, B_1^A(t,t^\prime) \Bigr] \;.
\end{split}
\end{align}
\end{subequations}

The boson-like boson-boson pair
\begin{equation}
  B(t,t^\prime) = B_1(t,t^\prime) \, B_2(t,t^\prime) \;,
\end{equation}
can be analytically continued by
\begin{subequations}
\begin{align}
  B^>(t,t^\prime) & = -i \, B_1^>(t,t^\prime) \, B_2^>(t,t^\prime) \;, \\
  B^<(t,t^\prime) & = -i \, B_1^<(t,t^\prime) \, B_2^<(t,t^\prime) \;, \\
\begin{split}
  B^R(t,t^\prime) & = -i \Bigl[ B_1^R(t,t^\prime) \, B_2^>(t,t^\prime) \\
& \phantom{= -i \Bigl[} + B_1^<(t,t^\prime) \, B_2^R(t,t^\prime) \Bigr] \;,
\end{split} \\
\begin{split}
  B^A(t,t^\prime) & = -i \Bigl[ B_1^A(t,t^\prime) \, B_2^>(t,t^\prime) \\
  & \phantom{= -i \Bigl[} + B_1^<(t,t^\prime) \, B_2^A(t,t^\prime) \Bigr] \;.
\end{split}
\end{align}
\end{subequations}

Finally, to analytically continue the boson-like boson-antiboson pair
\begin{equation}
  B(t,t^\prime) = B_1(t,t^\prime) \, B_2(t^\prime,t) \;,
\end{equation}
we use
\begin{subequations}
\begin{align}
  B^>(t,t^\prime) & = -i \, B_1^>(t,t^\prime) \, B_2^<(t^\prime,t) \;, \\
  B^<(t,t^\prime) & = -i \, B_1^<(t,t^\prime) \, B_2^>(t^\prime,t) \;, \\
\begin{split}
  B^R(t,t^\prime) & = -i \Bigl[ B_1^<(t,t^\prime) \, B_2^A(t^\prime,t) \\
& \phantom{= -i \Bigl[} + B_1^R(t,t^\prime) \, B_2^<(t^\prime,t) \Bigr] \;,
\end{split} \\
%\begin{split}
%  & = -i \Bigl[ B_1^>(t,t^\prime) \, B_2^A(t^\prime,t) \\
%  & \phantom{= -i \Bigl[} + B_1^R(t,t^\prime) \, B_2^>(t^\prime,t) \Bigr] \;,
%\end{split} \\
\begin{split}
  B^A(t,t^\prime) & = -i \Bigl[ B_1^<(t,t^\prime) \, B_2^R(t^\prime,t) \\
  & \phantom{= -i \Bigl[} + B_1^A(t,t^\prime) \, B_2^<(t^\prime,t) \Bigr] \;.
\end{split}% \\
%\begin{split}
%  & = -i \Bigl[ B_1^>(t,t^\prime) \, B_2^R(t^\prime,t) \\
%  & \phantom{= -i \Bigl[} + B_1^A(t,t^\prime) \, B_2^>(t^\prime,t) \Bigr] \;.
%\end{split}
\end{align}
\end{subequations}\\

Furthermore, for the analytic continuation of the contour integrals within the Dyson equations we also need to 
project terms of the form
\begin{equation}
  D(t,t^\prime) = \int_{\mathcal{C}} \mathrm{d} t_1 \; D_1(t,t_1) \, D_2(t_1,t^\prime) \;,
\end{equation}
where~$D$,~$D_1$ and~$D_2$ are either fermion-like or boson-like. This can be accomplished by the rules
\begin{subequations}
\begin{align}
\begin{split}
  D^<(t,t^\prime) & = \int_{-\infty}^\infty \mathrm{d}t_1 \; \Bigl[ D_1^R(t,t_1) \, D_2^<(t_1,t^\prime) \\
  & \phantom{= \int_{-\infty}^\infty \mathrm{d}t_1 \;} + D_1^<(t,t_1) \, D_2^A(t_1,t^\prime) \Bigr] \;,
\end{split} \\
\begin{split}
  D^>(t,t^\prime) & = \int_{-\infty}^\infty \mathrm{d}t_1 \; \Bigl[ D_1^R(t,t_1) \, D_2^>(t_1,t^\prime) \\
  & \phantom{= \int_{-\infty}^\infty \mathrm{d}t_1 \;} + D_1^>(t,t_1) \, D_2^A(t_1,t^\prime) \Bigr] \;,
\end{split} \\
  D^R(t,t^\prime) & = \int_{-\infty}^\infty \mathrm{d}t_1 \; D_1^R(t,t_1) \, D_2^R(t_1,t^\prime) \;, \\
  D^A(t,t^\prime) & = \int_{-\infty}^\infty \mathrm{d}t_1 \; D_1^A(t,t_1) \, D_2^A(t_1,t^\prime) \;.
\end{align}
\end{subequations}

\section{Dyson equations\label{DysonEq}}

In this appendix we summarize the Dyson equations for the analytic pieces of the molecular 
Green functions $G_-$, $B_*$, and $B_g$ as obtained by an application of the Langreth-Wilkins
rules of appendix \ref{app langreth} to the self-energies shown in Fig.~\ref{fig self energies}.

The lesser Green functions satisfy
\begin{widetext}
\begin{subequations}
\begin{align}
\begin{split}
  \left( i \frac{\partial}{\partial t} - \frac{\eps_-(t)}{\hbar} \right) G_-^<(t,t^\prime) = \int_{-\infty}^t\mathrm{d}t_1 \Bigl[ & \sigma_{\vec{k}}^>(t,t_1) \, B_*^R(t,t_1) \, G_-^<(t_1,t^\prime) + \sigma_{\vec{k}}^R(t,t_1) \, B_*^<(t,t_1) \, G_-^<(t_1,t^\prime) \\
  & + \sigma_{\vec{q}}^>(t,t_1) \, B_g^R(t,t_1) \, G_-^<(t_1,t^\prime) + \sigma_{\vec{q}}^R(t,t_1) \, B_g^<(t,t_1) \, G_-^<(t_1,t^\prime) \Bigr] \\
  + \int_{-\infty}^{t^\prime}\mathrm{d}t_1 \Bigl[ & \sigma_{\vec{k}}^<(t,t_1) \, B_*^<(t,t_1) \, G_-^A(t_1,t^\prime) + \sigma_{\vec{q}}^<(t,t_1) \, B_g^<(t,t_1) \, G_-^A(t_1,t^\prime) \Bigr] \;,
\end{split} \label{eq-plain-dyson-first} \tag{\theequation a} \\[1ex]
\begin{split}
  \left( i \frac{\partial}{\partial t} - \frac{\eps_{*}(t)}{\hbar} \right) B_{*}^<(t,t^\prime) = \int_{-\infty}^t\mathrm{d}t_1 \Bigl[ & \sigma_{\vec{k}}^<(t_1,t) \, G_-^R(t,t_1) \, B_{*}^<(t_1,t^\prime) + \sigma_{\vec{k}}^A(t_1,t) \, G_-^<(t,t_1) \, B_{*}^<(t_1,t^\prime) \\
  & + i \sigma_{\vec{k}\vec{q}}^>(t,t_1) \, B_g^R(t,t_1) \, B_{*}^<(t_1,t^\prime) + i \sigma_{\vec{k}\vec{q}}^R(t,t_1) \, B_g^<(t,t_1) \, B_{*}^<(t_1,t^\prime) \Bigr] \\
  + \int_{-\infty}^{t^\prime}\mathrm{d}t_1 \Bigl[ & \sigma_{\vec{k}}^>(t_1,t) \, G_-^<(t,t_1) \, B_{*}^A(t_1,t^\prime) + i \sigma_{\vec{k}\vec{q}}^<(t,t_1) \, B_g^<(t,t_1) \, B_{*}^A(t_1,t^\prime) \Bigr] \;,
\end{split} \tag{\theequation b} \\[1ex]
\begin{split}
  \left( i \frac{\partial}{\partial t} - \frac{\eps_{g}(t)}{\hbar} \right) B_{g}^<(t,t^\prime) = \int_{-\infty}^t\mathrm{d}t_1 \Bigl[ & \sigma_{\vec{q}}^<(t_1,t) \, G_-^R(t,t_1) \, B_{g}^<(t_1,t^\prime) + \sigma_{\vec{q}}^A(t_1,t) \, G_-^<(t,t_1) \, B_{g}^<(t_1,t^\prime) \\
  & + i \sigma_{\vec{k}\vec{q}}^<(t_1,t) \, B_*^R(t,t_1) \, B_{g}^<(t_1,t^\prime) + i \sigma_{\vec{k}\vec{q}}^A(t_1,t) \, B_*^<(t,t_1) \, B_{g}^<(t_1,t^\prime) \Bigr] \\
  + \int_{-\infty}^{t^\prime}\mathrm{d}t_1 \Bigl[ & \sigma_{\vec{q}}^>(t_1,t) \, G_-^<(t,t_1) \, B_{g}^A(t_1,t^\prime) + i \sigma_{\vec{k}\vec{q}}^>(t_1,t) \, B_*^<(t,t_1) \, B_{g}^A(t_1,t^\prime) \Bigr] \;,
\end{split} \tag{\theequation c}
\end{align}%
\end{subequations}
\addtocounter{equation}{-1}
while the retarded Green functions obey
\begin{subequations}
\begin{align}
\begin{split}
  \left( i \frac{\partial}{\partial t} - \frac{\eps_-(t)}{\hbar} \right) G_-^R(t,t^\prime) = \delta(t-t^\prime) + \int_{t^\prime}^t\mathrm{d}t_1 \Bigl[ & \sigma_{\vec{k}}^>(t,t_1) \, B_*^R(t,t_1) \, G_-^R(t_1,t^\prime) + \sigma_{\vec{k}}^R(t,t_1) \, B_*^<(t,t_1) \, G_-^R(t_1,t^\prime) \\
  & + \sigma_{\vec{q}}^>(t,t_1) \, B_g^R(t,t_1) \, G_-^R(t_1,t^\prime) + \sigma_{\vec{q}}^R(t,t_1) \, B_g^<(t,t_1) \, G_-^R(t_1,t^\prime) \Bigr] \;,
\end{split} \tag{\theequation d} \\[1ex]
\begin{split}
  \left( i \frac{\partial}{\partial t} - \frac{\eps_{*}(t)}{\hbar} \right) B_{*}^R(t,t^\prime) = \delta(t-t^\prime) + \int_{t^\prime}^t\mathrm{d}t_1 \Bigl[ & \sigma_{\vec{k}}^<(t_1,t) \, G_-^R(t,t_1) \, B_{*}^R(t_1,t^\prime) + \sigma_{\vec{k}}^A(t_1,t) \, G_-^<(t,t_1) \, B_{*}^R(t_1,t^\prime) \\
  & + i \sigma_{\vec{k}\vec{q}}^>(t,t_1) \, B_g^R(t,t_1) \, B_{*}^R(t_1,t^\prime) + i \sigma_{\vec{k}\vec{q}}^R(t,t_1) \, B_g^<(t,t_1) \, B_{*}^R(t_1,t^\prime) \Bigr] \;,\!\!\!\!\!
\end{split} \tag{\theequation e} \\[1ex]
\begin{split}
  \left( i \frac{\partial}{\partial t} - \frac{\eps_{g}(t)}{\hbar} \right) B_{g}^R(t,t^\prime) = \delta(t-t^\prime) + \int_{t^\prime}^t\mathrm{d}t_1 \Bigl[ & \sigma_{\vec{q}}^<(t_1,t) \, G_-^R(t,t_1) \, B_{g}^R(t_1,t^\prime) + \sigma_{\vec{q}}^A(t_1,t) \, G_-^<(t,t_1) \, B_{g}^R(t_1,t^\prime) \\
  & + i \sigma_{\vec{k}\vec{q}}^<(t_1,t) \, B_*^R(t,t_1) \, B_{g}^R(t_1,t^\prime) + i \sigma_{\vec{k}\vec{q}}^A(t_1,t) \, B_*^<(t,t_1) \, B_{g}^R(t_1,t^\prime) \Bigr] \;.\!\!\!\!\!
\end{split} \label{eq-plain-dyson-last} \tag{\theequation f}
\end{align}\label{eq-plain-dyson}%
\end{subequations}
The greater and advanced Green function can be obtained from the definitions 
\eqref{eq-retarded-definition} and the symmetry relations \eqref{SymmetryRelations}.\\

\end{widetext}

%\bibliography{main}
%\bibliographystyle{apsrev}

\end{document}